\documentclass[conference]{IEEEtran}
\IEEEoverridecommandlockouts

\usepackage[utf8]{inputenc}
\usepackage[caption=false, font=footnotesize]{subfig}
\usepackage{graphicx}
\usepackage{calc} 
\usepackage{cite}
\usepackage{amsmath,amssymb,amsfonts,amsthm}
\usepackage{algorithmic}
\usepackage{graphicx}
\usepackage{textcomp}
\usepackage{xcolor}
\usepackage{dirtytalk}
\usepackage{mathtools}

\DeclarePairedDelimiter\floor{\lfloor}{\rfloor}
\newtheorem{theorem}{Theorem}
\newtheorem{prop}{Proposition}
\newtheorem{corollary}{Corollary}[theorem]
\usepackage[caption=false, font=footnotesize]{subfig}
\usepackage{mdframed}
\usepackage{makecell}
\usepackage{caption}
\usepackage{float}
\usepackage{balance}
\newenvironment{theoencadre}
  {\begin{mdframed}\begin{theorem}}
  {\end{theorem}\end{mdframed}}
\newenvironment{coroencadre}
    {\begin{mdframed}\begin{corollary}}
    {\end{corollary}\end{mdframed}}
\newenvironment{propencadre}
    {\begin{mdframed}\begin{prop}}
    {\end{prop}\end{mdframed}}
\usepackage[euler]{textgreek}
\def\BibTeX{{\rm B\kern-.05em{\sc i\kern-.025em b}\kern-.08em
    T\kern-.1667em\lower.7ex\hbox{E}\kern-.125emX}}
\begin{document}

\title{On the Design of an Optimal Multi-Tone Jammer Against the Wiener Interpolation Filter\\
\thanks{This work has been supported in part by the project RIS3, under grant FRANCE 2030 ANR-23-CMAS-0023}
}

\author{\IEEEauthorblockN{Corentin Fonteneau}
\IEEEauthorblockA{\textit{Univ Rennes, INSA Rennes, IETR - UMR 6164 F-35000 Rennes, France} \\}
}

\maketitle

\begin{abstract} 

In the context of civilian and military communications, anti-jamming techniques are essential to ensure information integrity in the presence of malicious interference. A conventional time-domain approach relies on computing the Wiener interpolation filter to estimate and suppress the jamming waveform from the received samples. It is widely acknowledged that this method is effective for protecting wideband systems against narrowband interference.
In this work, this paradigm is questioned through the design of a $K$-tone jamming waveform that is intrinsically difficult to estimate assuming a $L$-tap Wiener interpolation filter. This design relies on an optimization procedure that maximizes the analytical Bayesian mean squared error associated with the jamming waveform estimate. Additionally, an analytical proof is provided showing that a multi-tone jamming waveform composed of $L/2+1$ tones is sufficient to render the Wiener-filter-based anti-jamming module completely ineffective. The analytical results are validated through Monte Carlo simulations assuming both perfect knowledge and practical estimates of the correlation functions of the received signal.


\end{abstract}


\section{Introduction}
Due to the continuous and growing digitalization of human activities, information security has become a critical concern aimed at maintaining control over services and infrastructures. In this context, modern systems must guarantee both the confidentiality and the integrity of the information to be transmitted \cite{liu_physical_2017,yue_low_2023,tusha_interference_2025,morales-ferre_survey_2020}. This paper focuses specifically on the latter, which characterizes the system’s ability to safeguard information from interference during transmission \cite{yue_low_2023}.


These interferences are commonly classified as unintentional — such as inter-carrier or co-channel interference — or intentional, when a malicious device disrupts the system \cite{tusha_interference_2025}. This intentional disruption can take the form of a spoofing attack, in which a counterfeit signal is transmitted to mislead the receiver, or a jamming attack \cite{morales-ferre_survey_2020}. More specifically, the latter consists in masking the useful signal by transmitting a strong interference within the frequency band of interest. Within this framework, multiple jamming waveforms, such as pulses or chirps, have been studied in the literature with the aim of enabling the decoding of the useful information in contested environments \cite{hegarty_suppression_2000,borio_gnss_2012,borio_swept_2016}.


In particular, the protection of wireless systems against narrow band interference (NBI)—often modeled as an autoregressive process of order 1 (AR(1)) or as a multi-tone signal—has attracted significant attention over the past decades \cite{aygur_narrowband_2025,masry_closed-form_1984,wang_rejection_1988,ketchum_adaptive_1982}. This type of interference has motivated the development of rejection techniques in the time \cite{ketchum_adaptive_1982}, frequency \cite{young_analysis_1998}, spatial \cite{krim_two_1996}, or space-time domains \cite{myrick_low_2001}. Regardless of the chosen domain, the underlying objective is to suppress the interference from the received signal in order to enhance the Signal-to-Interference-plus-Noise Ratio (SINR). To this end, various signal processing approaches, such as wavelet transforms \cite{musumeci_use_2014}, compressed sensing \cite{hu_narrowband_2025}, Robust Principal Component Analysis (RPCA) \cite{huang_reweighted_2019}, notch filtering \cite{borio_two-pole_2008}\cite{chien_design_2015}, and Wiener filtering \cite{masry_closed-form_1984}\cite{masry_closed-form_1985} have been investigated in different contexts.
     
More notably, the rejection of NBI using Wiener filtering in Direct-Sequence Spread Spectrum (DSSS) communications has been extensively discussed, as this architecture enables communication under high Jammer-to-Signal Ratio (JSR) conditions \cite{loh-ming_li_rejection_1982,ketchum_adaptive_1982,masry_closed-form_1984,iltis_performance_1984,masry_closed-form_1985,masry_performance_1986,wang_rejection_1988,fonteneau_rejection_2024}. In particular, the SINR is improved through both the filter gain and the inherent processing gain of the DSSS waveform. The two major advantages of the Wiener filter are, first, that it minimizes the Bayesian Mean Square Error (BMSE) of the interference estimate and, second, that it can be computed in a blind manner \cite{ketchum_adaptive_1982},\cite{Funda_Stat_Signal_Processing_EstTheory}. Subsequent works on Wiener filtering introduced a multi-stage architecture, known as the Multi-Stage Wiener Filter (MSWF), which achieves the same BMSE performance at a lower computational cost through rank reduction \cite{goldstein_multistage_1998}. Moreover, it requires fewer training samples to estimate the filter coefficients \cite{honig_adaptive_2002}. More recent studies have further enhanced the multi-stage framework by accelerating the convergence of the filter coefficients toward their optimal value and by improving algorithms used to select an appropriate dimension for the reduced-rank subspace \cite{song_adaptive_2012,zhang_fast_2016}.

The combination of Wiener filtering and DSSS systems has led to extensive analytical work and numerous simulation studies on the BMSE of the interference estimate for AR(1) and multi-tone jamming waveforms \cite{loh-ming_li_rejection_1982,ketchum_adaptive_1982,masry_closed-form_1984,masry_closed-form_1985,iltis_performance_1984,wang_rejection_1988,masry_performance_1986,fonteneau_rejection_2024}. Specifically, closed-form expressions for the analytical BMSE have been derived for BPSK-DSSS systems subjected to a sinusoidal jammer in \cite{loh-ming_li_rejection_1982,fonteneau_rejection_2024} and to an AR(1) jammer in \cite{masry_closed-form_1984,masry_closed-form_1985}. Furthermore, approximations of the analytical BMSE have been provided for both BPSK-DSSS and QPSK-DSSS systems under the assumption of a sum of AR(1) jammers or a multi-tone jammer in \cite{wang_rejection_1988}. Across these contributions, simulation results consistently indicate that substantial SINR improvements can be achieved through Wiener filtering, particularly in the presence of multi-tone jammers \cite{ketchum_adaptive_1982}. However, these studies rely on naive multi-tone jammer models, which is a strong assumption in scenarios involving a malicious interferer. To the best of the author's knowledge, the optimization of a narrowband jamming waveform has received limited attention \cite{masry_performance_1986,fonteneau_rejection_2024}. More precisely, the authors of \cite{masry_performance_1986} determined the constrained-bandwidth aliased spectrum of a jammer that maximizes the BMSE, assuming an infinite-length Wiener prediction filter. Their results show that the optimal aliased spectrum depends on the aliased spectra of both the useful signal and the noise at the receiver. Finally, analytical derivations in \cite{fonteneau_rejection_2024} established that, for a $L$-tap Wiener interpolation filter and a BPSK-DSSS system, the most difficult sinusoidal jammer to estimate is obtained for the normalized angular frequency $\omega_1=4.493/(L+1)$ or $\omega_1=\pi-4.493/(L+1)$.

\textbf{Contributions.} This paper addresses the design of a multi-tone jammer that is difficult to estimate using a Wiener interpolation filter in a DSSS system. The main contributions are summarized below:
\begin{itemize}
    \item A closed-form expression for the complex Wiener interpolation filter.
    \item A closed-form expression for the BMSE of the interference estimate.
    \item A closed-form expression for the optimal 2-tone jammer.
    \item A low-complexity procedure for generating the most difficult-to-estimate $K$-tone jammer for a $L$-tap filter.
    \item A closed-form characterization of the optimal $K$-tone jammer, assuming $K\geq L/2+1$. 
\end{itemize}

\textbf{Organization.} The paper is organized as follows. Section \ref{section : System Model} presents the system model. Section \ref{section : The Wiener Interpolation Filter} details the analytical performance of the anti-jamming module based on the Wiener interpolation filter. Section \ref{section : A systematic multi-tone jamming approach} discusses the design of an optimal multi-tone jamming waveform targeting this module. Section \ref{section : Simulation results} evaluates the relevance of the proposed jamming waveforms through Monte Carlo simulations. Finally, Section \ref{section : Conclusion} concludes the paper.

\textbf{Notations.} The following notations are used in the paper : $\textbf{A}$ is a matrix ; $\textbf{a}$ is a column vector ; $a$ is a scalar ; $\left[\mathbf{A}\right]_{ij}$ is the entry on the $i$th line and $j$th column of matrix $\mathbf{A}$; $\textbf{I}_N$ is an $N\times N$ identity matrix ;  $\textbf{1}_N$ is a $N$ all-ones column vector ; $\textbf{A}=\mathrm{diag}(a_1,...,a_{N})$ is a $N$x$N$ diagonal matrix with $a_i$ ($1\leq i\leq N$) elements on its diagonal diagonal matrix ; $\mathrm{Tr}(\mathbf{A})$ is the trace of $\mathbf{A}$ ; $(.)^*$, $(.)^T$, and  $(.)^H$ represent the conjugate, the transpose and the conjugate transpose respectively ; $\mathbb{E}$ is the expectation symbol.    

\section{System Model}
\label{section : System Model}
In this section, the received signal model is detailed in \ref{subsec : Received signal} before the presentation of the architecture of the anti-jamming module in \ref{subsec : General design of the anti-jamming module}. 
\subsection{Received signal}
\label{subsec : Received signal}

In this paper, a spread signal $s(t)$ is jammed by a multi-tone waveform $i(t)$ on a gaussian channel $n(t)$. The received signal $r(t)$ is sampled at chip rate under the assumption of perfect timing and frequency synchronization. 

In particular, the $m$th interference sample $i_m$ generated by the $K$-tone jammer is written,    
\begin{equation}
    \label{eq :i_m}
i_m(\boldsymbol{\alpha},\boldsymbol{\omega})=\mathbf{1}_K^T\boldsymbol{\theta}_m(\boldsymbol{\alpha},\boldsymbol{\omega}) \in \mathbb{C},
\end{equation}
where $\boldsymbol{\theta}_m(\boldsymbol{\alpha},\boldsymbol{\omega})\in\mathbb{C}^K$ is composed of the contribution of each individual tone. The contribution of the $k$th tone is expressed as,
\begin{equation}
\label{eqn : theta_m}
\left[\boldsymbol{\theta}_m(\boldsymbol{\alpha},\boldsymbol{\omega})\right]_k = \alpha_ke^{j(\omega_k m+\phi_k)}\in \mathbb{C},
\end{equation}
where $\alpha_k$, $\omega_k$ and $\phi_k$ respectively represent the modulus of the complex exponential, the normalized angular frequency in $[-\pi,\pi[$ and the random initial phase that follows a uniform distribution over $[-\pi,\pi[$ for the $k$-th tone. Note that $\boldsymbol{\alpha}=\left[\alpha_1,...,\alpha_K\right]^T\in(\mathbb{R}^+)^{K}$ and $\boldsymbol{\omega}=\left[\omega_1,...,\omega_K\right]^T\in[-\pi,\pi[^K$ are the jammer parameters that will be discussed and optimized in the rest of the paper.  

The $m$th received sample $r_m(\boldsymbol{\alpha},\boldsymbol{\omega})$ is then written,    
\begin{equation}
\label{eqn : r[n]}
r_m(\boldsymbol{\alpha},\boldsymbol{\omega})=s_m+n_m+i_m(\boldsymbol{\alpha},\boldsymbol{\omega}) \in \mathbb{C},
\end{equation}
where $s_m=\sqrt{S}\left(I_m+jQ_m\right)\in\mathbb{C}$ represents the $m$th useful sample of power $S$ ; $n_m$ is the $m$th complex additive white gaussian noise sample that follows the complex normal distribution $\mathcal{CN}\left(0,\sigma_n^2\right)$ and $i_m(\boldsymbol{\alpha},\boldsymbol{\omega})$ is the $m$th jamming sample arising from the $K$-tone jammer of power $J$. It is worth noting that $s_m$, $n_m$ and $i_m(\boldsymbol{\alpha},\boldsymbol{\omega})$ are mutually uncorrelated as well as Wide Sense Stationary (WSS) processes \cite{loh-ming_li_rejection_1982}. 

\subsection{General architecture of the anti-jamming module}
\label{subsec : General design of the anti-jamming module}

\begin{figure}[t!]
\begin{center}
\includegraphics[width=1\linewidth,trim={6cm 6cm 6cm 6cm},clip]{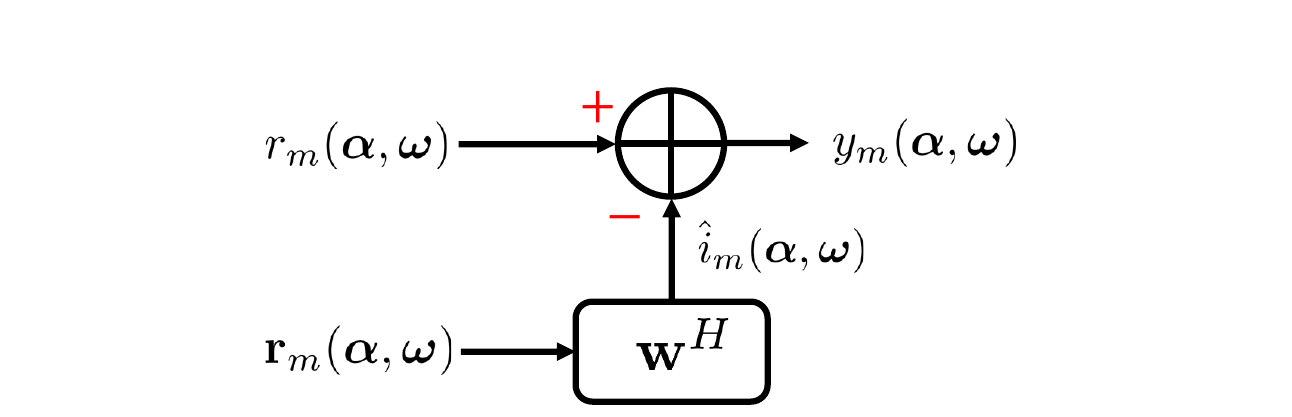}
\caption{Architecture of the anti-jamming module.}
\label{fig : Anti-jamming module architecture}
\end{center}
\end{figure}

 In order to enhance the robustness of the system in contested environments, the receiver is composed of an anti-jamming module depicted in Figure \ref{fig : Anti-jamming module architecture}. The estimation module produces an interference estimate $\hat{i}_m(\boldsymbol{\alpha},\boldsymbol{\omega})$ that is expressed as, 
 \begin{equation}
     \label{eqn : i_m est selon w}\hat{i}_m(\boldsymbol{\alpha},\boldsymbol{\omega})=\mathbf{w}^H\mathbf{r}_m(\boldsymbol{\alpha},\boldsymbol{\omega})\in\mathbb{C},
 \end{equation}
 which depends on the linear interpolation filter $\mathbf{w}\in\mathbb{C}^L$ of length $L$ and on the received samples $\mathbf{r}_m(\boldsymbol{\alpha},\boldsymbol{\omega})=\left[r_{m+L/2},\ldots,r_{m+1},r_{m-1},\ldots,r_{m-L/2}\right]^T\in\mathbb{C}^L$. 
 The estimate $\hat{i}_m(\boldsymbol{\alpha},\boldsymbol{\omega})$ is then subtracted from $r_m(\boldsymbol{\alpha},\boldsymbol{\omega})$ to obtain the output sample $y_m(\boldsymbol{\alpha},\boldsymbol{\omega})$ with improved SINR, which is written,
  \begin{equation}
     \label{eqn : y_m}y_m(\boldsymbol{\alpha},\boldsymbol{\omega})=r_m(\boldsymbol{\alpha},\boldsymbol{\omega})-\hat{i}_m(\boldsymbol{\alpha},\boldsymbol{\omega})\in\mathbb{C}.
 \end{equation}

At the receiver side, the main difficulty is to determine the filter $\mathbf{w}$ that produces an accurate interference estimate. A classical solution to this problem is to find the Linear Minimum Mean Square Error (LMMSE) estimator $\hat{i}^{\star}$ that minimizes the power $P_y$ of the output signal $y_m(\boldsymbol{\alpha},\boldsymbol{\omega})$, i.e  \cite{loh-ming_li_rejection_1982},
\begin{align}
\label{eqn : opt problem w}
    \mathbf{w}^{\star} =\underset{\mathbf{w}}{\arg \min}\;  \mathbb{E}_{s,n,\phi}\left[|y_m(\boldsymbol{\alpha},\boldsymbol{\omega})|^2\right].
\end{align}
The resolution of such a problem is equivalent to the minimization of the BMSE of the interference estimate since,
\begin{align}
\label{eqn : P_y}
    P_y &= \mathbb{E}_{s,n,\phi}\left[\big|s_m+n_m+\left(i_m(\boldsymbol{\alpha},\boldsymbol{\omega})-\hat{i}_m(\boldsymbol{\alpha},\boldsymbol{\omega})\right)\big|^2\right],\nonumber\\
    &=S+\sigma_n^2+\mathrm{BMSE}\big(\,\hat{i}(\boldsymbol{\alpha},\boldsymbol{\omega})\,\big),
\end{align}
where,
\begin{equation}
    \label{eqn : BMSE i est}\mathrm{BMSE}\big(\,\hat{i}(\boldsymbol{\alpha},\boldsymbol{\omega})\,\big)= \mathbb{E}_{s,n,\phi}\left[|i_m(\boldsymbol{\alpha},\boldsymbol{\omega})-\hat{i}_m(\boldsymbol{\alpha},\boldsymbol{\omega})|^2\right].
\end{equation}
When the processes are zero mean and WSS, as in the present case, the solution $\mathbf{w}^{\star}$ to this problem is commonly referred to as the Wiener filter \cite{Funda_Stat_Signal_Processing_EstTheory}. The analytical form of $\mathbf{w}^{\star}$ and the BMSE of the interference estimate $\hat{i}^{\star}(\boldsymbol{\alpha},\boldsymbol{\omega})$ are the subject of next section.  

\section{Analytical performance of the anti-jamming module}
\label{section : The Wiener Interpolation Filter}
In this section, the analytical performance of Wiener filtering in DSSS systems is studied in depth. Firstly, the complex vector form of the Wiener interpolation filter $\mathbf{w}^{\star}$ is analytically derived in section \ref{section : Analytical solution of the Wiener filter for a multi-tone jammer in DSSS systems}.
Then, the analytical BMSE of the interference estimate $\hat{i}^{\star}(\boldsymbol{\alpha},\boldsymbol{\omega})$ is established in section \ref{section : Analytical BMSE  for a multi-tone jammer in DSSS systems}. Finally, the blind estimation of $\mathbf{w}^{\star}$ is discussed in section \ref{section : Blind estimation of the Wiener interpolation filter}.  
\subsection{Analytical solution of the Wiener interpolation filter $\mathbf{w}^{\star}$}
\label{section : Analytical solution of the Wiener filter for a multi-tone jammer in DSSS systems}
As expressed by \eqref{eq :i_m}, the jamming sample $i_m(\boldsymbol{\alpha},\boldsymbol{\omega})$ is a sum of $K$ unknown jamming samples $\left[\boldsymbol{\theta}_m(\boldsymbol{\alpha},\boldsymbol{\omega})\right]_k$ given by \eqref{eqn : theta_m}. Reminding that the LMMSE estimator of a sum of unknown parameters is the sum of the individual estimators, the interference estimate $\hat{i}_m^{\star}(\boldsymbol{\alpha},\boldsymbol{\omega})$ is written \cite[P390]{Funda_Stat_Signal_Processing_EstTheory},
\begin{equation}
    \label{eqn : i_m est selon theta}\hat{i}_m^{\star}(\boldsymbol{\alpha},\boldsymbol{\omega})=\mathbf{1}_K^T\hat{\boldsymbol{\theta}}_m^{\star}(\boldsymbol{\alpha},\boldsymbol{\omega}),
\end{equation}
where $\hat{\boldsymbol{\theta}}_m^{\star}(\boldsymbol{\alpha},\boldsymbol{\omega})=\left[\hat{\theta}_m^{\star}{(\alpha_1,\omega_1)},\ldots,\hat{\theta}_m^{\star}{(\alpha_K,\omega_K)}\right]^T\in\mathbb{C}^K$ consists of one estimate per tone. As samples are zero mean and WSS, $\hat{\boldsymbol{\theta}}_m^{\star}(\boldsymbol{\alpha},\boldsymbol{\omega})$ is written \cite[eq.(12.20)]{Funda_Stat_Signal_Processing_EstTheory},
\begin{align}
    \label{eqn : theta_m est}\hat{\boldsymbol{\theta}}_m^{\star}(\boldsymbol{\alpha},\boldsymbol{\omega})= \mathbf{C}_{\mathbf{r}\boldsymbol{\theta}}^H(\boldsymbol{\alpha},\boldsymbol{\omega})\mathbf{C}_{\mathbf{rr}}^{-1}(\boldsymbol{\alpha},\boldsymbol{\omega})\mathbf{r}_m(\boldsymbol{\alpha},\boldsymbol{\omega}),
\end{align}
where $\mathbf{C}_{\mathbf{rr}}(\boldsymbol{\alpha},\boldsymbol{\omega})\in\mathbb{C}^{L\times L}$ and $\mathbf{C}_{\mathbf{r}\boldsymbol{\theta}}(\boldsymbol{\alpha},\boldsymbol{\omega})\in\mathbb{C}^{L\times K}$ respectively denote the covariance matrix and the cross-covariance matrix for $L$ observations and $K$ unknown parameters.  
From \eqref{eqn : i_m est selon w}, \eqref{eqn : i_m est selon theta} and \eqref{eqn : theta_m est}, the analytical expression of the complex Wiener interpolation filter $\mathbf{w}^{\star}$ can be determined in the special case of a multi-tone jammer in DSSS systems. This is the subject of Theorem \ref{th : wiener filter}: 
\pagebreak
\begin{theoencadre}
\label{th : wiener filter}
The Wiener interpolation filter $\mathbf{w}^{\star}$ for a multi-tone jammer in DSSS systems is expressed as, 
\begin{align}
\label{eqn : general form of the wiener filter}
\mathbf{w}^{\star}=\mathbf{C}_{\mathbf{r}\mathbf{r}}^{-1}(\boldsymbol{\alpha},\boldsymbol{\omega})\mathbf{C}_{\mathbf{r}\boldsymbol{\theta}}(\boldsymbol{\alpha},\boldsymbol{\omega})\boldsymbol{1}_K\in\mathbb{C}^K,
\end{align}
where, 
\begin{equation}
\label{eqn : C_rr}
    \mathbf{C}_{\mathbf{r}\mathbf{r}}(\boldsymbol{\alpha},\boldsymbol{\omega})=(S+\sigma_n^2)\mathbf{I}_L+\boldsymbol{\Psi}(\boldsymbol{\omega})\mathbf{J}(\boldsymbol{\alpha})\boldsymbol{\Psi}^H(\boldsymbol{\omega})\in\mathbb{C}^{L\times L},
\end{equation}
\begin{equation}
    \label{eqn : C_rtheta}\mathbf{C}_{\mathbf{r}\boldsymbol{\theta}}(\boldsymbol{\alpha},\boldsymbol{\omega})=\boldsymbol{\Psi}(\boldsymbol{\omega})\mathbf{J}(\boldsymbol{\alpha})\in\mathbb{C}^{L\times K},
\end{equation}
\begin{equation}
\label{eqn : J_matrix}
\mathbf{J}(\boldsymbol{\alpha})=\mathrm{diag}\left(\alpha_1^2,\alpha_2^2,\ldots,\alpha_K^2\right)\in(\mathbb{R}^+)^{K \times K} ,
\end{equation}
\begin{equation}
\boldsymbol{\Psi}(\boldsymbol{\omega})=\left[\boldsymbol{\psi}({\omega_1}),\boldsymbol{\psi}({\omega_2}),\ldots,\boldsymbol{\psi}({\omega_K})\right]\in\mathbb{C}^{L\times K},
\end{equation}
\begin{equation}
\label{eqn : psi_k}
\boldsymbol{\psi}({\omega_k})=\left[e^{j\frac{L}{2}\omega_k},\ldots,e^{j\omega_k},e^{-j\omega_k},\ldots,e^{-j\frac{L}{2}\omega_k}\right]^T\in\mathbb{C}^{L}.
\end{equation}
\begin{proof}
Appendix \ref{appendix : Wiener Filter}
\end{proof}
\end{theoencadre}
As an example, the estimation of a 4-tone jammer is qualitatively studied, assuming perfect knowledge of the matrices $\mathbf{C}_{\mathbf{r}\mathbf{r}}(\boldsymbol{\alpha},\boldsymbol{\omega})$ and $\mathbf{C}_{\mathbf{r}\boldsymbol{\theta}}(\boldsymbol{\alpha},\boldsymbol{\omega})$ in a QPSK-DSSS system characterized by a Signal-to-Noise Ratio (SNR) of $-15\, \mathrm{dB}$ and a JSR of $25\, \mathrm{dB}$. 

\begin{figure*}[!t]
    \centering
    \subfloat[Estimation of the real part of the interference for $L=64$.]{%
        \includegraphics[width=0.48\textwidth]{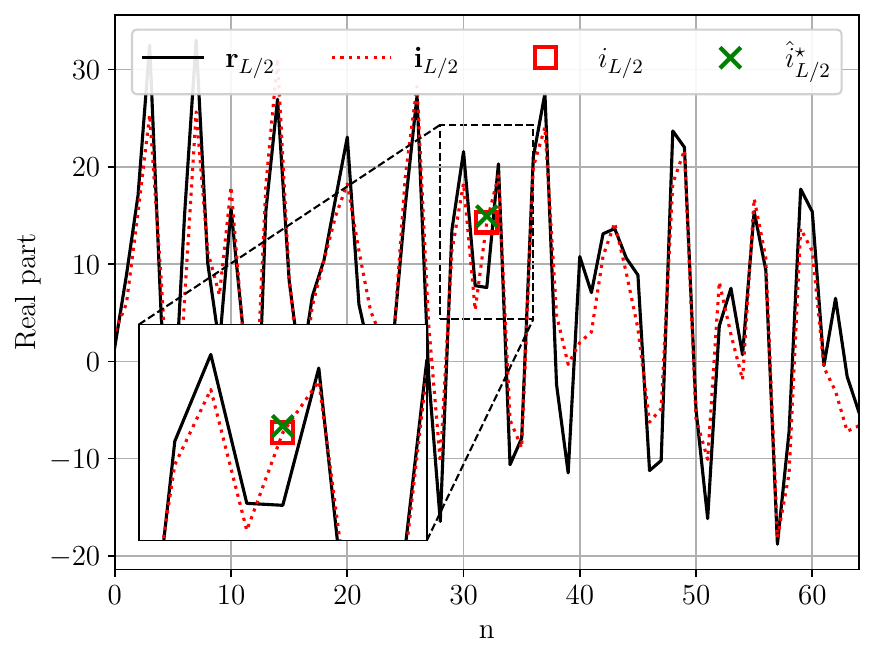}%
        \label{fig:top_left}
    }\hfill
    \subfloat[Estimation of the imaginary part of the interference for $L=64$.]{%
        \includegraphics[width=0.48\textwidth]{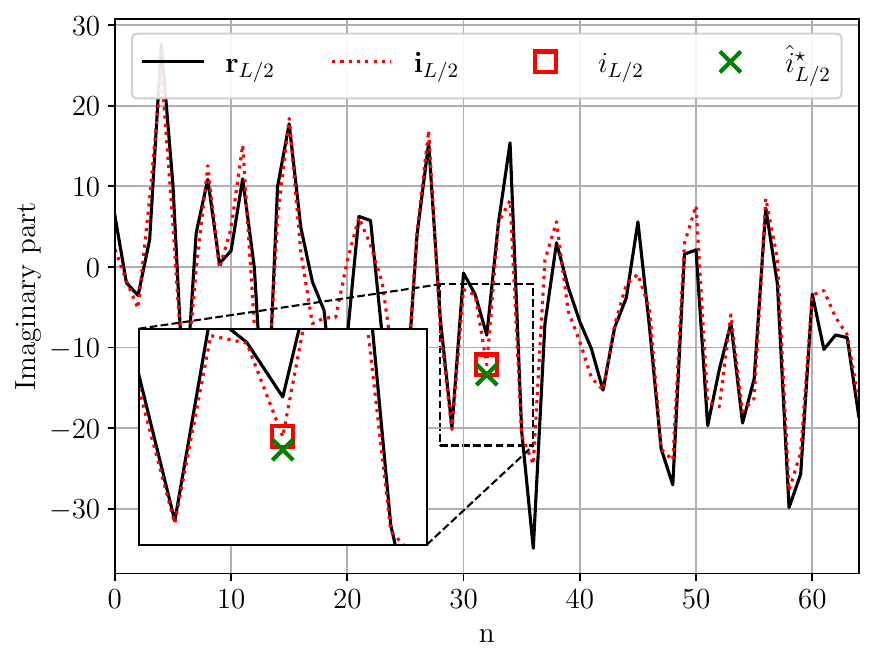}%
        \label{fig:top_right}
    }
\vspace{0.5cm}
\subfloat[Estimation of the individual interfering tones $\boldsymbol{\theta}_{L/2}$ and of the resulting jamming sample $i_{L/2}$, assuming various random draws of $\mathbf{s}_{L/2}$ and $\mathbf{n}_{L/2}$ for a fixed scenario $\{\boldsymbol{\alpha},\boldsymbol{\omega},\boldsymbol{\phi}\}$.]{%
        \includegraphics[width=0.48\textwidth]{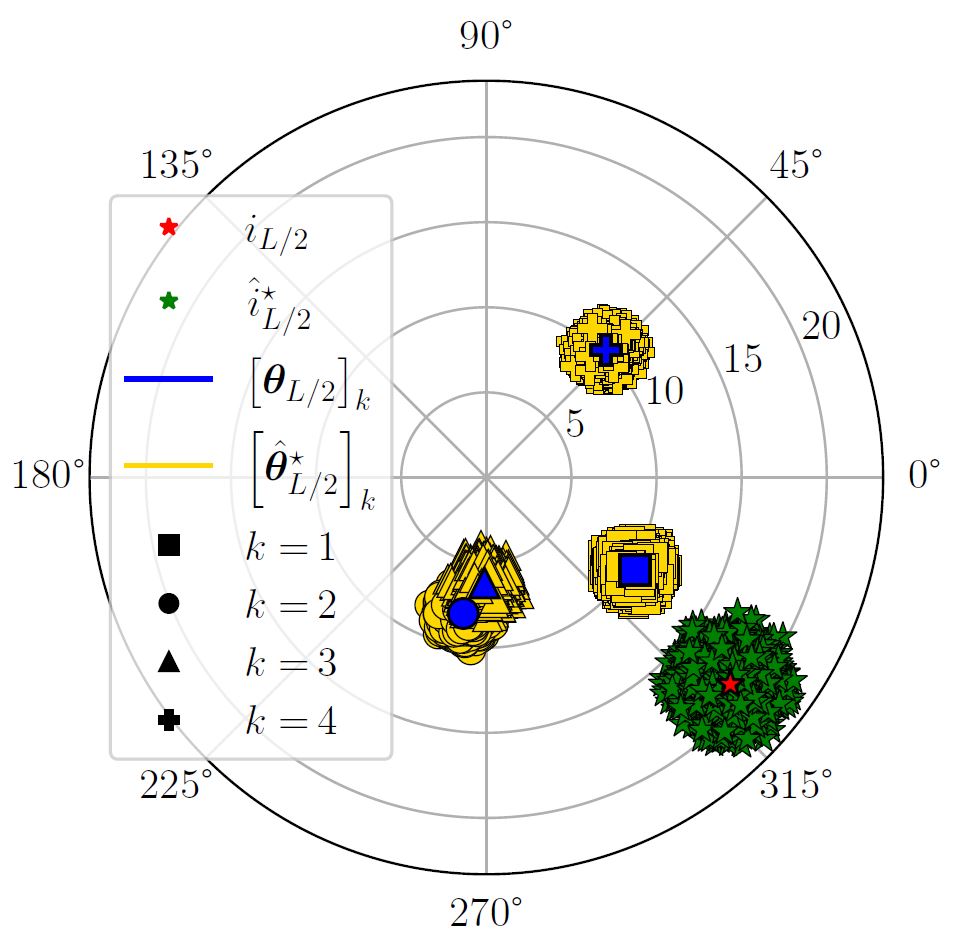}%
        \label{fig:bottom_left}
    }\hfill
    \subfloat[PDF of the module and phase of the estimated interference $\hat{i}_{L/2}^{\star}$ with respect to the filter length $L$ of the Wiener interpolation, assuming various random draws of $\mathbf{s}_{L/2}$ and $\mathbf{n}_{L/2}$ for a fixed scenario $\{\boldsymbol{\alpha},\boldsymbol{\omega},\boldsymbol{\phi}\}$.]{%
        \includegraphics[width=0.48\textwidth]{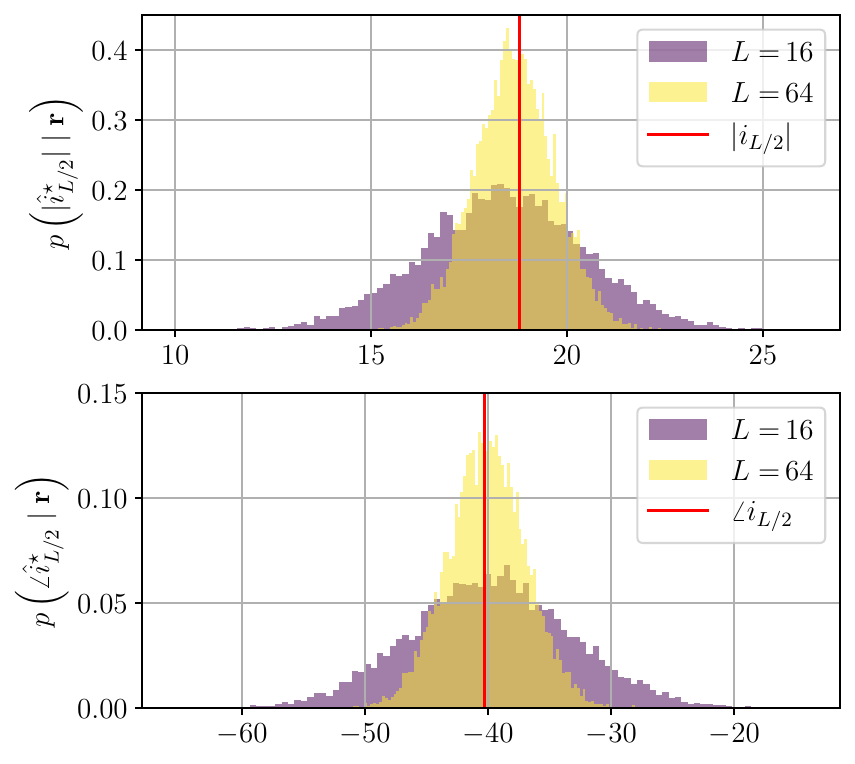}%
        \label{fig:bottom_right}
    }
    \caption{Estimation of a 4-tone jammer in a QPSK-DSSS system by Wiener filtering ($SNR=-15\, \mathrm{dB}$, $JSR=25\, \mathrm{dB}$).}
    \label{fig:estim_4toneJammer_WienerFilter}
\end{figure*}

Specifically, Figures \ref{fig:top_left} and \ref{fig:top_right} show that the interference estimate $\hat{i}_{L/2}^{\star}(\boldsymbol{\alpha},\boldsymbol{\omega})$, obtained by applying \eqref{eqn : i_m est selon w} and \eqref{eqn : general form of the wiener filter}, closely matches both the real part and imaginary parts of the complex jamming sample $i_{L/2}(\boldsymbol{\alpha},\boldsymbol{\omega})$ for a filter length $L=64$. Additionally, Figure \ref{fig:bottom_left} depicts the interference estimate $\left[\hat{\boldsymbol{\theta}}_{L/2}^{\star}(\boldsymbol{\alpha},\boldsymbol{\omega})\right]_k$ for each tone, as well as the resulting multi-tone estimate $\hat{i}_{L/2}^{\star}(\boldsymbol{\alpha},\boldsymbol{\omega})$, assuming  a fixed scenario $\{\boldsymbol{\alpha},\boldsymbol{\omega},\boldsymbol{\phi}\}$ over multiple random realizations of the useful signal $\mathbf{s}_{L/2}$ and the additive white gaussian noise $\mathbf{n}_{L/2}$. This illustration highlights that, even with perfect knowledge of the matrices $\mathbf{C}_{\mathbf{r}\mathbf{r}}(\boldsymbol{\alpha},\boldsymbol{\omega})$ and $\mathbf{C}_{\mathbf{r}\boldsymbol{\theta}}(\boldsymbol{\alpha},\boldsymbol{\omega})$, the error in the interference estimate $\hat{i}_{L/2}^{\star}(\boldsymbol{\alpha},\boldsymbol{\omega})$ depends on the specific realizations of $\mathbf{s}_{L/2}$ and $\mathbf{n}_{L/2}$. Finally, the Figure \ref{fig:bottom_right} depicts that the Probability Density Function (PDF) of the interference estimate is sensitive to the length $L$ of the Wiener interpolation filter, i.e, to the number of samples that are available for interference estimation. As shown in the figure, the higher the value of $L$, the lower the dispersion and the better the quality of the estimate $\hat{i}_{L/2}^{\star}(\boldsymbol{\alpha},\boldsymbol{\omega})$. 

This preliminary qualitative analysis on the estimation error is complemented in section \ref{section : Analytical BMSE  for a multi-tone jammer in DSSS systems} by the analytical derivation of the BMSE of the interference estimate $\hat{i}^{\star}(\boldsymbol{\alpha},\boldsymbol{\omega})$, obtained for a fixed jamming scenario $\{\boldsymbol{\alpha},\boldsymbol{\omega}\}$ under random realizations $\{\boldsymbol{\phi},\mathbf{s},\mathbf{n}\}$. 

\subsection{Analytical BMSE of the interference estimate $\hat{i}^{\star}(\boldsymbol{\alpha},\boldsymbol{\omega})$}
\label{section : Analytical BMSE  for a multi-tone jammer in DSSS systems}

The BMSE of the interference estimate $\hat{i}^{\star}(\boldsymbol{\alpha},\boldsymbol{\omega})$ is a key metric for the interference suppression problem, as it quantifies the quality of the interference estimation and, consequently, the suppression capability of the anti-jamming module. This is the subject of Theorem \ref{th : analytical BMSE}:
\begin{theoencadre}
\label{th : analytical BMSE}
The BMSE of the interference estimate $\hat{i}^{\star}(\boldsymbol{\alpha},\boldsymbol{\omega})$ for a multi-tone jammer in DSSS systems is expressed as,
\begin{equation}
\label{eqn : BMSE i est star}
    \mathrm{BMSE}\big(\,\hat{i}^{\star}(\boldsymbol{\alpha},\boldsymbol{\omega})\,\big)= \boldsymbol{1}_K^T\mathbf{C}_{\boldsymbol{\epsilon}}^{\star}(\boldsymbol{\alpha},\boldsymbol{\omega})\boldsymbol{1}_K,
\end{equation}
with,
\begin{equation}
\label{eqn : C_espsilon star}
    \mathbf{C}_{\boldsymbol{\epsilon}}^{\star}(\boldsymbol{\alpha},\boldsymbol{\omega})=\mathbf{A}^{-1}(\boldsymbol{\alpha},\boldsymbol{\omega})\mathbf{J}(\boldsymbol{\alpha})\in\mathbb{R}^{K\times K},
\end{equation}
\begin{equation}
\label{eqn : A matrix}
    \mathbf{A}(\boldsymbol{\alpha},\boldsymbol{\omega})=\mathbf{I}_K+(S+\sigma_n^2)^{-1}\mathbf{J}(\boldsymbol{\alpha})\boldsymbol{\Gamma}(\boldsymbol{\omega})\in\mathbb{R}^{K\times K},
\end{equation}
\begin{equation}
    \boldsymbol{\Gamma}(\boldsymbol{\omega})=\boldsymbol{\Psi}^H(\boldsymbol{\omega})\boldsymbol{\Psi}(\boldsymbol{\omega})\in\mathbb{R}^{K\times K}
\end{equation}
\begin{equation}
\label{eqn : Gamma_ij}
    \left[\boldsymbol{\Gamma}(\boldsymbol{\omega})\right]_{kk'}=D_{L/2}\left(\Delta_{k'k}\right)-1\in\mathbb{R},
\end{equation}
where $\mathbf{C}_{\boldsymbol{\epsilon}}^{\star}(\boldsymbol{\alpha},\boldsymbol{\omega})$ is the minimum error covariance matrix, $\Delta_{k'k}=w_{k'}-w_k$ is the normalized angular frequency difference between the $k'$th tone and the $k$th one, and $D_{n}(x)=\sin{\left((n+1/2)x\right)}/\sin{\left(x/2\right)}$ denotes the $n$th Dirichlet kernel.
\begin{proof}
    Appendix \ref{eqn : bayesian MSE}
\end{proof}
\end{theoencadre}
This theorem highlights that the BMSE of the interference estimate $\hat{i}^{\star}(\boldsymbol{\alpha},\boldsymbol{\omega})$ depends not only on the BMSEs of individual tone estimates, i.e $\left[\mathbf{C}_{\boldsymbol{\epsilon}}^{\star}(\boldsymbol{\alpha},\boldsymbol{\omega})\right]_{kk}=\mathrm{BMSE}(\hat{\theta}^{\star}{(\alpha_k,\omega_k)})$, but also on the correlation of the estimation errors, i.e $\left[\mathbf{C}_{\boldsymbol{\epsilon}}^{\star}(\boldsymbol{\alpha},\boldsymbol{\omega})\right]_{kk^{'}},k\neq k^{'}$. The design method introduced in Section \ref{section : A systematic multi-tone jamming approach} exploits this property to generate a multi-tone jamming waveform that is difficult to estimate.

\subsection{Blind estimation of the Wiener interpolation filter $\mathbf{w}^{\star}$}
\label{section : Blind estimation of the Wiener interpolation filter}
Before discussing the design of the multi-tone jammer in the next section, the blind estimation of the Wiener interpolation filter $\mathbf{w}^{\star}$ is addressed as this approach is exploited in Section \ref{section : Simulation results} to obtain the numerical results.   

First, it's worth noting that the term $\mathbf{C}_{\mathbf{r}\boldsymbol{\theta}}(\boldsymbol{\alpha},\boldsymbol{\omega})\boldsymbol{1}_K$ \eqref{eqn : general form of the wiener filter} can be rewritten solely in terms of the received signal. Indeed, 
\begin{align}
    \mathbf{C}_{\mathbf{r}\boldsymbol{\theta}}(\boldsymbol{\alpha},\boldsymbol{\omega})\boldsymbol{1}_K&=\mathbb{E}_{s,n,\phi}\left[\mathbf{r}_m(\boldsymbol{\alpha},\boldsymbol{\omega})\boldsymbol{\theta}_m^H(\boldsymbol{\alpha},\boldsymbol{\omega})\right]\boldsymbol{1}_K,\nonumber\\
    &=\mathbb{E}_{s,n,\phi}\left[\mathbf{r}_m(\boldsymbol{\alpha},\boldsymbol{\omega})i_m^*(\boldsymbol{\alpha},\boldsymbol{\omega})\right],\nonumber\\
    &=\mathbf{c}_{\mathbf{r}i}(\boldsymbol{\alpha},\boldsymbol{\omega}),\nonumber\\
    &=\mathbb{E}_{s,n,\phi}\left[\mathbf{r}_m(\boldsymbol{\alpha},\boldsymbol{\omega},\boldsymbol{\phi})r_m^*(\boldsymbol{\alpha},\boldsymbol{\omega})\right].
\end{align}
Consequently, the matrix $\mathbf{C}_{\mathbf{r}\mathbf{r}}(\boldsymbol{\alpha},\boldsymbol{\omega})$ and  the vector $\mathbf{c}_{\mathbf{r}i}(\boldsymbol{\alpha},\boldsymbol{\omega})$ can be directly estimated from the received signal \cite{ketchum_adaptive_1982,zhang_fast_2016},
\begin{equation}
\label{eqn : C_rr est}
    \hat{\mathbf{C}}_{\mathbf{r}\mathbf{r}}(\boldsymbol{\alpha},\boldsymbol{\omega})=\frac{1}{M}\sum_{m=L/2}^{L/2+M-1}{\mathbf{r}_m(\boldsymbol{\alpha},\boldsymbol{\omega})\mathbf{r}_m^H(\boldsymbol{\alpha},\boldsymbol{\omega})},
\end{equation}
\begin{equation}
\label{eqn : c_ri est}
    \hat{\mathbf{c}}_{\mathbf{r}i}(\boldsymbol{\alpha},\boldsymbol{\omega})=\frac{1}{M}\sum_{m=L/2}^{L/2+M-1}{\mathbf{r}_m(\boldsymbol{\alpha},\boldsymbol{\omega})r_m^*(\boldsymbol{\alpha},\boldsymbol{\omega})},
\end{equation}
where $\hat{\mathbf{C}}_{\mathbf{r}\mathbf{r}}(\boldsymbol{\alpha},\boldsymbol{\omega})$, $\hat{\mathbf{c}}_{\mathbf{r}i}(\boldsymbol{\alpha},\boldsymbol{\omega})$ and $M$ respectively denote the estimate of the covariance matrix of the received samples, the estimate of the cross-covariance vector between the received samples and the interference, and the number of $L$-size vector used for the computation. Once \eqref{eqn : C_rr est} and \eqref{eqn : c_ri est} are computed, $\mathbf{w}^{\star}$ is obtained by multiplying the inverse of $\hat{\mathbf{C}}_{\mathbf{r}\mathbf{r}}(\boldsymbol{\alpha},\boldsymbol{\omega})$ to $\hat{\mathbf{c}}_{\mathbf{r}i}(\boldsymbol{\alpha},\boldsymbol{\omega})$.

\section{Design of an Optimal Multi-Tone Jammer}
\label{section : A systematic multi-tone jamming approach}
In this section, the design of an optimal multi-tone jammer against the Wiener interpolation filter is investigated. First, the optimization problem is formulated in section \ref{section : Optimization problem}. The problem is then analyzed and solved for $K=1$ tone in section \ref{section : Optimization for K=1 tone}, for $K=2$ tones in section \ref{section : Optimization for K=2 tones}, for $3\leq K<\frac{L}2{+1}$ tones in section \ref{section : Optimization for 3 to L/2+1 tones}, and for $K\geq\frac{L}2{+1}$ in section \ref{section : K geq to L/2 + 1}.        
\subsection{Optimization problem}
\label{section : Optimization problem}
Assuming that the receiver embeds an anti-jamming module based on the Wiener interpolation filter $\mathbf{w}^{\star}$, the jammer's goal is to maximize the BMSE of the interfence estimate $\hat{i}^{\star}(\boldsymbol{\alpha},\boldsymbol{\omega})$ \eqref{eqn : BMSE i est star}. Recalling that the jammer power at the receiver is $J$, the optimization problem can be formulated as,
\begin{align}
\label{eqn : optimization problem max BMSE}
    &\underset{\boldsymbol{\alpha},\,\boldsymbol{\omega}}{\textrm{minimize}}\; f(\boldsymbol{\alpha},\boldsymbol{\omega}) = - \mathrm{BMSE}\big(\,\hat{i}^{\star}(\boldsymbol{\alpha},\boldsymbol{\omega})\,\big),
    \\
    &\textrm{subject to } \| \boldsymbol{\alpha}\|_2^2 = J\nonumber
\end{align}
where $f(\boldsymbol{\alpha},\boldsymbol{\omega})$ denotes the loss function. The following subsections discuss the minimization of this function with respect to the number of jamming tones $K$.


\subsection{Optimization for $K=1$ tone}
\label{section : Optimization for K=1 tone}
For a jamming waveform composed of $K=1$ tone, the modulus $\alpha_1$ equals $\sqrt{J}$ to satisfy the power constraint. The BMSE of the interference estimate $\hat{i}^{\star}(\sqrt{J},\omega_1)$ is the subject of Corollary \ref{prop : 1 tone jammer}: 
\begin{coroencadre}
\label{prop : 1 tone jammer}
The BMSE of the interference estimate $\hat{i}^{\star}(\sqrt{J},\omega_1)$ for a 1-tone jamming waveform is expressed as,
\begin{equation}
\label{eq : BMSE 1 tone}
    \mathrm{BMSE}\big(\,\hat{i}^{\star}(\sqrt{J},{\omega_1})\,\big)=\frac{S+\sigma_n^2}{\frac{S+\sigma_n^2}{J}+L}.
\end{equation}
\begin{proof}
    Noticing that $\boldsymbol{\Gamma}(\omega_1)=L$ and $\mathbf{J}(\sqrt{J})=J$, \eqref{eq : BMSE 1 tone} is directly obtained from \eqref{eqn : BMSE i est star} to \eqref{eqn : A matrix}.
\end{proof}
\end{coroencadre}
From Corollary \ref{prop : 1 tone jammer}, it is figured out that the BMSE of the interference estimate $\hat{i}^{\star}(\sqrt{J},{\omega_1})$ does not depend on the optimization variable $\omega_1$. Consequently, the interference estimate can only be degraded by increasing the jammer power $J$.       

\subsection{Optimization for $K=2$ tones}
\label{section : Optimization for K=2 tones}

In order to determine the optimal 2-tone jammer, the original problem \eqref{eqn : optimization problem max BMSE} is reformulated as follows,  
\begin{align}
\label{eqn : optimization problem Lagrangian}
    \underset{\boldsymbol{\alpha},\,\boldsymbol{\omega},\lambda}{\textrm{minimize}}\; \mathcal{L}(\boldsymbol{\alpha},\boldsymbol{\omega},\lambda) = f(\boldsymbol{\alpha},\boldsymbol{\omega})+\lambda\left(-J+\sum_{k=1}^{K}{\alpha_k^2}\right),
\end{align}
where $\mathcal{L}(\boldsymbol{\alpha},\boldsymbol{\omega},\lambda)$ and $\lambda$ respectively denote the Lagrangian function and the Lagrange multiplier. To solve \eqref{eqn : optimization problem Lagrangian}, the partial derivatives of $\mathcal{L}(\boldsymbol{\alpha},\boldsymbol{\omega},\lambda)$ with respect to $\omega_k$ and $\alpha_k^2$ are first calculated regardless the value of $K$. This is the subject of Theorem \ref{th : partial derivatives}:

\begin{theoencadre}
\label{th : partial derivatives}
The partial derivative of the Lagrangian function $\mathcal{L}(\boldsymbol{\alpha},\boldsymbol{\omega},\lambda)$ with respect to $\alpha_k^2$ is,
\begin{equation}
\label{eqn : partial derivative of BMSE according to alpha_i^2}
    \frac{\partial \mathcal{L}(\boldsymbol{\alpha},\boldsymbol{\omega},\lambda)}{\partial \alpha_k^2}=-\left(\boldsymbol{1}_K^T\mathbf{e}_k\right)^2+\lambda,
\end{equation}
\begin{align}
\label{eqn : E matrix}
    \mathbf{E}(\boldsymbol{\alpha},\boldsymbol{\omega})=\left[\mathbf{e}_1,\ldots,\mathbf{e}_k,\ldots,\mathbf{e}_K\right]=\mathbf{A}^{-1}(\boldsymbol{\alpha},\boldsymbol{\omega}),
\end{align}
and the partial derivative of the Lagrangian function $\mathcal{L}(\boldsymbol{\alpha},\boldsymbol{\omega},\lambda)$ with respect to $\omega_k$ is,
\begin{align}
\label{eqn : partial derivative of the loss function with respect to omega_i}
    \frac{\partial \mathcal{L}(\boldsymbol{\alpha},\boldsymbol{\omega},\lambda)}{\partial \omega_k}=\frac{\boldsymbol{1}_K^T}{S+\sigma_n^2}\mathbf{C}_{\boldsymbol{\epsilon}}(\boldsymbol{\alpha},\boldsymbol{\omega})\frac{\partial \boldsymbol{\Gamma}(\boldsymbol{\omega})}{\partial \omega_k}\mathbf{C}_{\boldsymbol{\epsilon}}(\boldsymbol{\alpha},\boldsymbol{\omega})\boldsymbol{1}_K,
\end{align}
\small
\begin{align}
\label{eqn : partial derivative of gamma matrix with respect to omega_i}
    \left[\frac{\partial \boldsymbol{\Gamma}(\boldsymbol{\omega})}{\partial \omega_k}\right]_{kk^{'}}=\left[\frac{\partial \boldsymbol{\Gamma}(\boldsymbol{\omega})}{\partial \omega_k}\right]_{k^{'}k}=\left\{ 
    \begin{array}{l}
\frac{\partial D_{L/2}\left(\Delta_{k^{'}k}\right)}{\partial \omega_k}, k\neq k^{'}
\\
0 \mathrm{\;elsewhere}
    \end{array}\right. ,
\end{align}
\small
\begin{align}
\label{eqn : partial derivative of the Dirichlet kernel}
&\frac{\partial D_{L/2}\left(\Delta_{k^{'}k}\right)}{\partial \omega_k}=\left[\frac{1}{2}\sin{\left((L+1)\frac{\Delta_{k^{'}k}}{2}\right)}\cos{\left(\frac{\Delta_{k^{'}k}}{2}\right)}-\right.\nonumber\\
&\left.\frac{L+1}{2}\cos{\left((L+1)\frac{\Delta_{k^{'}k}}{2}\right)}\sin{\left(\frac{\Delta_{k^{'}k}}{2}\right)}\right]/\sin^2{\left(\frac{\Delta_{k^{'}k}}{2}\right)}.
\end{align}

\begin{proof}
    Appendix \ref{Calculation of partial derivatives}
\end{proof}

\end{theoencadre}
From partial derivatives ${\partial \mathcal{L}(\boldsymbol{\alpha},\boldsymbol{\omega},\lambda)}/{\partial \alpha_k^2}$ and ${\partial \mathcal{L}(\boldsymbol{\alpha},\boldsymbol{\omega},\lambda)}/{\partial \omega_k}$, the convexity of the Lagrangian function $\mathcal{L}(\boldsymbol{\alpha},\boldsymbol{\omega},\lambda)$ is studied for $K=2$ and the optimal solution with regard to $L$ is determined. This is the subject of Proposition \ref{Prop : 2-tone jamming signal}: 


\begin{propencadre}
\label{Prop : 2-tone jamming signal}
The 2-tone jammer that maximizes the BMSE of the interference estimate is parameterized as follows,
\begin{equation}
    \label{eqn : alpha dagger}\boldsymbol{\alpha}^{\dagger}=\left[\sqrt{J/2},\sqrt{J/2}\right]^T,
\end{equation}
\begin{equation}
    \label{eqn : omega dagger}\boldsymbol{\omega}^{\dagger}=\left[\omega_1,\omega_1+\Delta_{21}^{\dagger}\right]^T,
\end{equation}
\begin{equation}
\label{eqn : Delta_21 dagger}
    \Delta_{21}^{\dagger}=\left\{ 
    \begin{array}{l}
            9/(L+1) \\
            2\pi-9/(L+1)
    \end{array}
    \right. .
\end{equation}
The BMSE obtained for this jamming waveform is expressed as,
\begin{equation}
    \label{eqn : BMSE 2 tons}\mathrm{BMSE}\left(\,\hat{i}^{\star}\left(\boldsymbol{\alpha}^{\dagger},\boldsymbol{\omega}^{\dagger}\right)\,\right)=\frac{S+\sigma_n^2}{\frac{S+\sigma_n^2}{J}+\frac{L-1}{2}+\frac{D_{L/2}(\Delta_{21}^{\dagger})}{2}}.
\end{equation}
\begin{proof}
Appendix \ref{2-tone jamming signal optimization}
\end{proof}
\end{propencadre}

First, it is pointed out from \eqref{eqn : alpha dagger} to \eqref{eqn : Delta_21 dagger} that there is an infinite number of optimal 2-tone jammers rather than a single one. These jammers share the common characteristics of being composed of two equal-power tones, whose normalized angular frequency difference equals $\Delta_{21}^{\dagger}$. Consequently, in accordance with \eqref{eqn : BMSE 2 tons}, the absolute frequency positions of these tones have no impact on the BMSE of the interference estimate $\hat{i}^{\star}\left(\boldsymbol{\alpha}^{\dagger},\boldsymbol{\omega}^{\dagger}\right)$. 
Finally, it is worth noting from \eqref{eq : BMSE 1 tone} and \eqref{eqn : BMSE 2 tons} that, regardless of $\Delta_{21}$, a jammer composed of two equal-power tones is intrinsically harder to estimate than its 1-tone counterpart as $D_{L/2}(\Delta_{21})\leq L+1$.

\subsection{Optimization for $ 3\leq K<\frac{L}2{+1}$ tones}
\label{section : Optimization for 3 to L/2+1 tones}
The resolution of problem \eqref{eqn : optimization problem max BMSE} requires the inversion of matrix $\mathbf{A}(\boldsymbol{\alpha},\boldsymbol{\omega})\in\mathbb{R}^{K\times K}$, thereby complicating the analytical derivation of the optimal jammer for $3\leq K<\frac{L}2{+1}$. Therefore, a gradient-based optimization algorithm is employed to iteratively compute $\boldsymbol{\alpha}$ and $\boldsymbol{\omega}$. The general update rule at the $i$th iteration is given by,    
\begin{align}
\label{eqn : updating rule} 
    \left\{ 
    \begin{array}{l}
    \boldsymbol{\alpha}_{(i)} =\mathcal{P}_{\mathcal{S}}(\boldsymbol{\alpha}_{(i-1)}-\mu_{\alpha}\nabla_{\boldsymbol{\alpha}}f(\boldsymbol{\alpha}_{(i-1)},\boldsymbol{\omega}_{(i-1)}))\\
    \boldsymbol{\omega}_{(i)} =\boldsymbol{\omega}_{(i-1)}-\mu_{\omega}\nabla_{\boldsymbol{\omega}}f(\boldsymbol{\alpha}_{(i-1)},\boldsymbol{\omega}_{(i-1)})
    \end{array}
    \right. ,
\end{align}
where the projection operator $\mathcal{P}_{\mathcal{S}}(\boldsymbol{\alpha})=\sqrt{J}\times\boldsymbol{\alpha}/\|\boldsymbol{\alpha}\|_2$ ensures that the power constraint is satisfied after each iteration, and $\mu_{\alpha}$ (resp. $\mu_{\omega}$) denotes the step size associated with $\boldsymbol{\alpha}$ (resp. $\boldsymbol{\omega}$). In practice, the optimization process is initialized using multiple pairs of random initial vectors $\{\boldsymbol{\alpha}_{(0)}^{\mathrm{r}},\boldsymbol{\omega}_{(0)}^{\mathrm{r}}\}$ and the resulting solution $\{\boldsymbol{\alpha}_{\mathrm{opt}}^{\mathrm{r}},\boldsymbol{\omega}_{\mathrm{opt}}^{\mathrm{r}}\}$ that minimizes $f(\boldsymbol{\alpha},\boldsymbol{\omega})$ at convergence is selected to generate the jamming waveform. Nevertheless, this method can be computationally relaxed by properly selecting the initial pair $\{\boldsymbol{\alpha}_{(0)},\boldsymbol{\omega}_{(0)}\}$. 

To this end, it is first noted that the proposition \ref{Prop : 2-tone jamming signal} gives the optimal difference $\Delta_{21}^{\dagger}$ that jointly maximizes the BMSE of individual tone estimates, i.e $\left[\mathbf{C}_{\boldsymbol{\epsilon}}^{\star}(\boldsymbol{\alpha}^{\dagger},\boldsymbol{\omega}^{\dagger})\right]_{11}=\mathrm{BMSE}(\hat{\theta}^{\star}{(\alpha_1^{\dagger},\omega_1^{\dagger})})$ and $\left[\mathbf{C}_{\boldsymbol{\epsilon}}^{\star}(\boldsymbol{\alpha}^{\dagger},\boldsymbol{\omega}^{\dagger})\right]_{22}=\mathrm{BMSE}(\hat{\theta}^{\star}{(\alpha_2,\omega_2)})$, as well as the correlation of the estimation errors, i.e $\left[\mathbf{C}_{\boldsymbol{\epsilon}}^{\star}(\boldsymbol{\alpha}^{\dagger},\boldsymbol{\omega}^{\dagger})\right]_{12}$ and $\left[\mathbf{C}_{\boldsymbol{\epsilon}}^{\star}(\boldsymbol{\alpha}^{\dagger},\boldsymbol{\omega}^{\dagger})\right]_{21}$. Following the same waveform design philosophy, $\boldsymbol{\alpha}_{(0)}$ and $\boldsymbol{\omega}_{(0)}$ are chosen such that the jamming tones are equally powered and spaced by $\Delta_{21}^{\dagger}$ to penalize both the BMSE of individual tone estimates $\left[\mathbf{C}_{\boldsymbol{\epsilon}}^{\star}(\boldsymbol{\alpha}^{\dagger},\boldsymbol{\omega}^{\dagger})\right]_{kk}$ and the correlation of estimation errors linked to adjacent tones, i.e $\left[\mathbf{C}_{\boldsymbol{\epsilon}}^{\star}(\boldsymbol{\alpha}^{\dagger},\boldsymbol{\omega}^{\dagger})\right]_{kk-1}$ and $\left[\mathbf{C}_{\boldsymbol{\epsilon}}^{\star}(\boldsymbol{\alpha}^{\dagger},\boldsymbol{\omega}^{\dagger})\right]_{kk+1}$. This leads to the following parameterization,
\begin{equation}
    \label{eqn : alpha dagger 3 to L/2+1}\boldsymbol{\alpha}_{(0)}^{\dagger}=\sqrt{J/K}\times\boldsymbol{1}_K^T,
\end{equation}
\begin{equation}
    \label{eqn : omega dagger 3 to L/2+1}\left[\boldsymbol{\omega}_{(0)}^{\dagger}\right]_k=\omega_1+(k-1)\times\Delta_{21}^{\dagger},
\end{equation}
 with $\Delta_{21}^{\dagger}=9/(L+1)$, where the initial pair $\{\boldsymbol{\alpha}_{(0)}^{\dagger},\boldsymbol{\omega}_{(0)}^{\dagger}\}$ is expected to be in the vicinity of the global minimum of the function $f(\boldsymbol{\alpha},\boldsymbol{\omega})$. However, this solution ignores the fact that a correlation exists between estimation errors linked to non-adjacent tones. To address this issue, it is proposed to refine the jamming waveform by solely updating the vector $\boldsymbol{\omega}_{(i)}^{\dagger}$ with a per-tone correction term $\Delta_k^{\mathrm{opt}}$, which is obtained via gradient-based optimization. This leads to the solution,
\begin{equation}
    \label{eqn : omega dagger opt 3 to L/2+1}\left[\boldsymbol{\omega}_{\mathrm{opt}}^{\dagger}\right]_k=\left[\boldsymbol{\omega}_{(0)}^{\dagger}\right]_k+\Delta_k^{\mathrm{opt}},
\end{equation}
where $\Delta_k^{\mathrm{opt}}$ denotes the correction term associated with the $k$th tone. In the rest of the paper, the Adam optimizer is used to obtain $\{\boldsymbol{\alpha}_{\mathrm{opt}}^{\mathrm{r}},\boldsymbol{\omega}_{\mathrm{opt}}^{\mathrm{r}}\}$ and $\boldsymbol{\omega}_{\mathrm{opt}}^{\dagger}$ \cite{kingma_adam_2017}. The initial learning rates $\mu_{\alpha}$ and $\mu_{\omega}$ are set to $5\times10^{-2}$ and $10^{-2}$ respectively, and the maximum number of iterations is parameterized to $10^{3}$. 

\begin{figure}[t!]
\begin{center}
\includegraphics[width=1\linewidth,trim={0cm 0cm 0cm 0cm},clip]{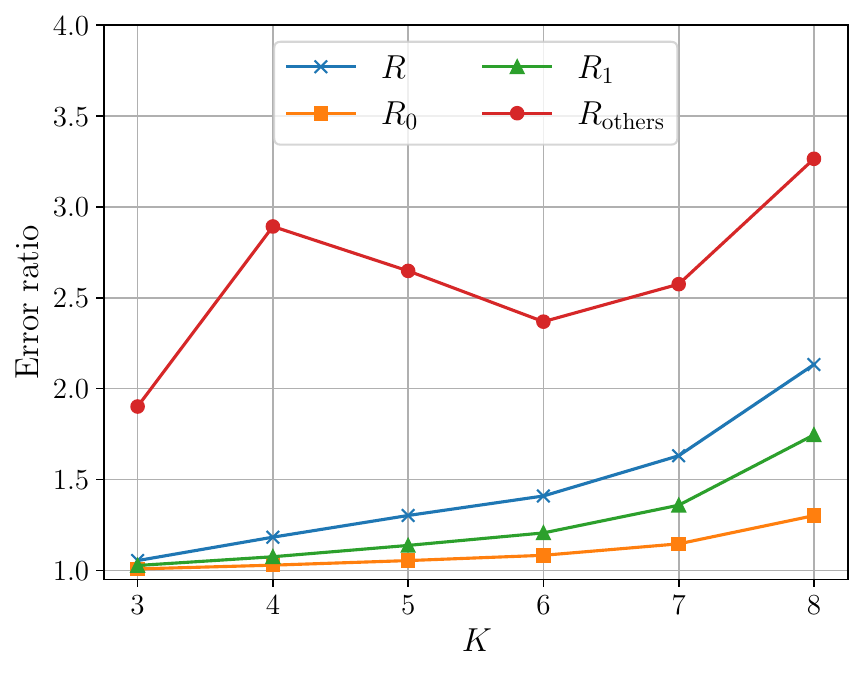}
\caption{Increase in error ratio obtained with the help of the optimizer.}
\label{fig : increase in error ratio}
\end{center}
\end{figure}

Firstly, the hypothesis that the correction term $\Delta_k^{\mathrm{opt}}$ mainly increases the correlation of estimation errors linked to non-adjacent tones is assessed in Figure \ref{fig : increase in error ratio} for jamming waveforms composed of $K=3$ to $K=8$ tones, assuming a $16$-tap filter. The  following metrics are introduced for this evaluation,      
\begin{equation}
    \label{eqn : R}R=\frac{N}{D}=\frac{\mathrm{BMSE}\big(\,\hat{i}^{\star}(\boldsymbol{\alpha}_{(0)}^{\dagger},\boldsymbol{\omega}^{\dagger}_{\mathrm{opt}})\,\big)}{\mathrm{BMSE}\big(\,\hat{i}^{\star}(\boldsymbol{\alpha}_{(0)}^{\dagger},\boldsymbol{\omega}_{(0)}^{\dagger})\,\big)},
\end{equation}
\begin{equation}
\label{eqn : R0}
    R_0=\frac{N_0}{D_0}=\frac{\mathrm{Tr}(\mathbf{C}_{\boldsymbol{\epsilon}}^{\star}(\boldsymbol{\alpha}_{(0)}^{\dagger},\boldsymbol{\omega}_{\mathrm{opt}}^{\dagger}))}{\mathrm{Tr}(\mathbf{C}_{\boldsymbol{\epsilon}}^{\star}(\boldsymbol{\alpha}_{(0)}^{\dagger},\boldsymbol{\omega}_{(0)}^{\dagger}))},
\end{equation}
\begin{equation}
\label{eqn : R1}
    R_1=\frac{N_1}{D_1}=\frac{\sum_{k=1}^{K-1}{\left[\mathbf{C}_{\boldsymbol{\epsilon}}^{\star}(\boldsymbol{\alpha}_{(0)}^{\dagger},\boldsymbol{\omega}_{\mathrm{opt}}^{\dagger})\right]_{k(k+1)}}}{\sum_{k=1}^{K-1}{\left[\mathbf{C}_{\boldsymbol{\epsilon}}^{\star}(\boldsymbol{\alpha}_{(0)}^{\dagger},\boldsymbol{\omega}_{(0)}^{\dagger})\right]_{k(k+1)}}},
\end{equation}
\begin{equation}
    \label{eqn : R_others}R_{\mathrm{others}}=\frac{N-N_0-2N_1}{D-D_0-2D_1},
\end{equation}
where $R$, $R0$, $R_1$ and $R_{\mathrm{others}}$ respectively denote the increase in error ratio obtained with the help of the optimizer for the BMSE of the interference estimate $\hat{i}^{\star}$ \eqref{eqn : R}, for the BMSE of individual tone estimates $\hat{\theta}_k^{\star}$ \eqref{eqn : R0}, for the correlation of estimation errors related to adjacent tones \eqref{eqn : R1} and for the correlation of estimation errors related to non-adjacent tones \eqref{eqn : R_others}. As observed in Figure \ref{fig : increase in error ratio}, the error ratio $R_{\mathrm{others}}$ is greatly improved by the optimization process, which confirms the relevance of the additional term $\Delta_k^{\mathrm{opt}}$ to increase the correlation of estimation errors related to non-adjacent tones. More moderate gains are obtained for $R_0$ and $R_1$, which attests that the initial point $\{\boldsymbol{\alpha}_{(0)}^{\dagger},\boldsymbol{\omega}_{(0)}^{\dagger}\}$ is an efficient way of penalizing both the BMSE of individual tone estimates $\hat{\theta}_k^{\star}$ and the correlation of estimation errors linked to adjacent tones. 

\begin{figure}[t!]
\begin{center}
\includegraphics[width=1\linewidth,trim={0cm 0cm 0cm 0cm},clip]{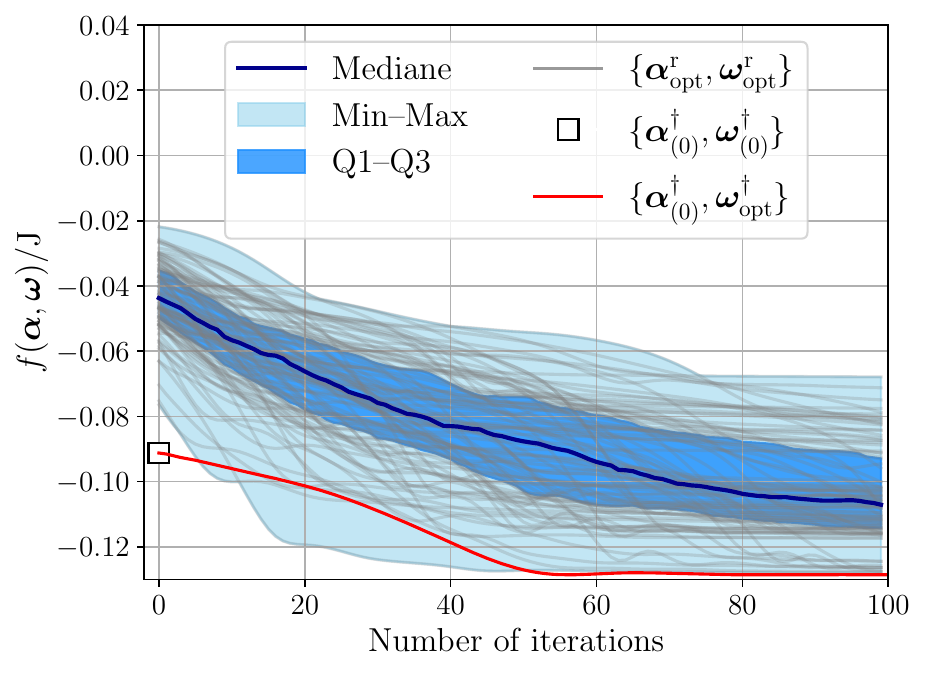}
\caption{Jamming waveform optimisation for $K=6$ tones and $L=16$.}
\label{fig:Jamming waveform optimisation for K=6}
\end{center}
\end{figure}

Finally, the optimization process of solutions $\{\boldsymbol{\alpha}_{\mathrm{opt}}^{\mathrm{r}},\boldsymbol{\omega}_{\mathrm{opt}}^{\mathrm{r}}\}$ and $\{\boldsymbol{\alpha}_{(0)}^{\dagger},\boldsymbol{\omega}_{\mathrm{opt}}^{\dagger}\}$ is depicted in Figure \ref{fig:Jamming waveform optimisation for K=6}, in which the normalized objective function $f(\boldsymbol{\alpha},\boldsymbol{\omega})$ is represented with respect to the number of iterations of the optimizer. Initial vectors $\{\boldsymbol{\alpha}_{(0)}^{\mathrm{r}},\boldsymbol{\omega}_{(0)}^{\mathrm{r}}\}$ are randomly drawn from uniform distributions. As observed in this figure, the optimization process associated to $\{\boldsymbol{\alpha}_{(0)}^{\dagger},\boldsymbol{\omega}_{\mathrm{opt}}^{\dagger}\}$ has the advantage of converging faster to the global minima and makes optimization of multiple initial points useless, which is desirable from a complexity point of view.       

\subsection{Optimization for $K\geq\frac{L}2{+1}$ tones}
\label{section : K geq to L/2 + 1}
No more optimization process is needed for $K\geq\frac{L}2{+1}$ tones. This is the subject of Proposition \ref{Prop : (L/2+1)-tone jamming signal}:
\pagebreak
\begin{propencadre}
\label{Prop : (L/2+1)-tone jamming signal}
A K-tone jammer which is parameterized as follows,
\begin{equation}
    \label{eqn : alpha vector K L/2 + 1 tones}\boldsymbol{\alpha}^{\ddagger}=\sqrt{J/K}\times\boldsymbol{1}_K^T,
\end{equation}
\begin{equation}
    \label{eqn : omega vector K L/2 + 1 tones}\left[\boldsymbol{\omega}^{\ddagger}\right]_k=\omega_1+(k-1)\times\frac{2\pi}{K},
\end{equation}
leads to the following BMSE for the interference estimate  $\hat{i}^{\star}\left(\boldsymbol{\alpha}^{\ddagger},\boldsymbol{\omega}^{\ddagger}\right)$,
\begin{equation}
    \label{eqn : BMSE ddagger}\mathrm{BMSE}\left(\,\hat{i}^{\star}\left(\boldsymbol{\alpha}^{\ddagger},\boldsymbol{\omega}^{\ddagger}\right)\,\right)=\left[J^{-1}+\frac{2}{S+\sigma_n^2} \floor*{\frac{L}{2K}}\right]^{-1}. 
\end{equation}
This parameterization becomes optimal for $K\geq\frac{L}2{+1}$ tones as $\mathrm{BMSE}\left(\,\hat{i}^{\star}\left(\boldsymbol{\alpha}^{\ddagger},\boldsymbol{\omega}^{\ddagger}\right)\,\right)=J$.
\begin{proof}
Appendix \ref{appendix : BMSE calculation for the optimal jammer composed of K=(L/2)+1 tones}
\end{proof}
\end{propencadre}
First, it is pointed out from \eqref{eqn : BMSE ddagger} that increasing the number of tones does not always mean improving the jammer performance. Indeed, two $K$-tone jammers with equal ratio $\floor*{\frac{L}{2K}}$ will perform the same. Additionally, in contested environments for which $J>>(S+\sigma_n^2)$ holds, it becomes almost ineffective to further increase the jammer power $J$ in order to enlarge the BMSE when $K<L/2+1$. This behavior is explained by the fact that the BMSE in \eqref{eqn : BMSE ddagger} is dominated by the leading term $\frac{2}{(S+\sigma_n^2)}\floor*{\frac{L}{2K}}$ in the denominator, making the jamming waveform poorly effective. On the contrary, the jamming waveform becomes highly effective when $K\geq\frac{L}2{+1}$ tones. In this case, the BMSE equals $J$, indicating that this jamming waveform constitutes an optimal solution to \eqref{eqn : optimization problem max BMSE} as the anti-jamming module no longer provides any suppression capability. Last but not least, this result highlights that a jammer can choose $K$ in order to make all the anti-jamming modules based on a Wiener interpolation filter of length $L\leq 2(K-1)$ completely ineffective.         
\section{Simulation results}
\label{section : Simulation results}
\begin{figure}[t!]
\begin{center}
\includegraphics[width=1\linewidth,trim={0cm 0cm 0cm 0cm},clip]{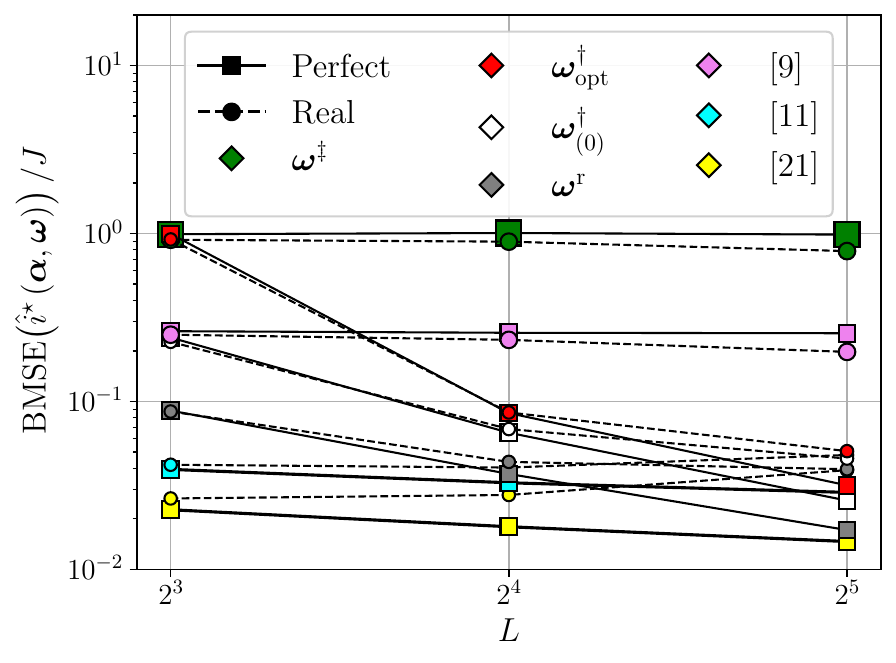}
\caption{Empirical BMSE of the interference estimate ($SNR=-15\, \mathrm{dB}$, $JSR=25\, \mathrm{dB}$).}
\label{fig : Empirical BMSE with the estimator i hat}
\end{center}
\end{figure}

In this section, the efficiency of the suggested jamming waveform design is assessed by performing a Monte-Carlo simulation for a receiver embedding a Wiener interpolation filter $\mathbf{w}^{\star}$ of length $L\in\{8,16,32\}$ in a QPSK-DSSS scenario. The normalized BMSE of the interference estimate $\hat{i}^{\star}(\boldsymbol{\alpha},\boldsymbol{\omega})$ is depicted in Figure \ref{fig : Empirical BMSE with the estimator i hat} for various jamming waveforms, which are detailed in the next paragraph. Whatever the jamming waveform, the filter $\mathbf{w}^{\star}$ is computed assuming either a perfect knowledge of $\mathbf{C}_{\mathbf{r}\mathbf{r}}(\boldsymbol{\alpha},\boldsymbol{\omega})$ \eqref{eqn : C_rr} and $\mathbf{C}_{\mathbf{r}\boldsymbol{\theta}}(\boldsymbol{\alpha},\boldsymbol{\omega})$ \eqref{eqn : C_rtheta} or the real estimates $\hat{\mathbf{C}}_{\mathbf{r}\mathbf{r}}(\boldsymbol{\alpha},\boldsymbol{\omega})$ \eqref{eqn : C_rr est} and $\hat{\mathbf{c}}_{\mathbf{r}i}(\boldsymbol{\alpha},\boldsymbol{\omega})$ \eqref{eqn : c_ri est}. In the latter case, these estimates are computed in a limited time window of $N=128$ samples, with $M=N-L$. This choice has been made as it is important to restrict the value $N$ in practice to ensure that the anti-jamming module can adapt to possible changes of the jamming waveform \cite{ketchum_adaptive_1982}. 

Six equal-power tones jamming waveforms are compared. Three of them come from the jamming waveform design suggested in section \ref{section : A systematic multi-tone jamming approach}: a 5-tone jammer parameterized by $\boldsymbol{\omega}_{(0)}^{\dagger}$ \eqref{eqn : omega dagger 3 to L/2+1}, a 5-tone jammer parameterized by $\boldsymbol{\omega}_{\mathrm{opt}}^{\dagger}$ \eqref{eqn : omega dagger opt 3 to L/2+1} and a $({L}/{2}+1)$-tone jammer parameterized by $\boldsymbol{\omega}^{\ddagger}$ \eqref{eqn : omega vector K L/2 + 1 tones}. The three others directly come from the literature : an equally spaced 100-tone jammer spanning 20\% of the bandwidth \cite{ketchum_adaptive_1982}, an equally spaced 10-tone jammer with an interval of 3.6 degrees between adjacent tones \cite{loh-ming_li_rejection_1982} and a 5-tone jammer $\boldsymbol{\omega}^{\mathrm{r}}$ for which the angular frequencies are randomly drawn from an uniform distribution between $-\pi$ and $\pi$. Note that the latter approach aims to represent articles for which the chosen angular frequencies are given without being discussed \cite{chien_design_2015}. Additionally, a seventh jamming waveform, based on an AR(1) process, is considered as this approach has been shown to be more effective than multiband jamming \cite{masry_closed-form_1984,masry_closed-form_1985}. The covariance sequence of this jamming waveform is given by $\rho_k=J\times \alpha^{|k|}$ and $\alpha=0.8$ has been chosen to be in line with \cite{masry_closed-form_1984}. 

First, the performance of jammers is analyzed assuming perfect knowledge of correlation functions. It is observed that the jamming approaches considered in \cite{ketchum_adaptive_1982} and \cite{loh-ming_li_rejection_1982} are poorly effective as less than 4\% of the jammer power remains after the anti-jamming module. The random jammer $\boldsymbol{\omega}^{\mathrm{r}}$ composed of 5 tones is slightly more effective than the latter approaches for $L=8$ but becomes less and less effective as $L$ increases. This result indicates that, when $L$ is high compared to $K$, it is more effective to jam a portion of the bandwidth with a sufficient density of tones, as suggested by \cite{ketchum_adaptive_1982} and \cite{loh-ming_li_rejection_1982}, rather than random frequency components. As expected, the AR(1) jammer is much more effective with a residual power that equals 30\% of the jammer power after the anti-jamming module. Interestingly, this last approach is matched by the 5-tone jammer with the analytical parameterization $\boldsymbol{\omega}_{(0)}^{\dagger}$ and surpassed by the parameterization $\boldsymbol{\omega}_{\mathrm{opt}}^{\dagger}$ obtained after optimization for $L=8$. Note that the residual power obtained with the latter reaches 100\%  of the jammer power, indicating that the anti-jamming module becomes completely ineffective. However, the jammer efficiency of such approaches decreases significantly as $L$ increases since the number of tones $K$ is insufficient. This analysis is confirmed by the performance of the jammer parameterized by $\boldsymbol{\omega}^{\ddagger}$ which ensures the ineffectiveness of the anti-jamming-module regardless of $L$ by being defined over a sufficiently large number of tones, namely $K=L/2+1$.

Second, the performance of jammers is studied assuming real estimates of correlation functions. With the exception of $\boldsymbol{\omega}^{\ddagger}$, it is observed that the BMSE is slightly higher than the one obtained with the perfect knowledge counterpart whatever the jamming parameterization. This result indicates that the quality of the interference estimation is degraded by the partial knowledge of the correlation functions, which in turn reduces the performance of the anti-jamming module. Conversely, partial knowledge of these functions can be beneficial from the receiver's perspective when the jammer is parameterized by $\boldsymbol{\omega}^{\ddagger}$, since this jamming approach is optimal under perfect knowledge of the correlation functions.  

These results highlight that the jamming waveform design detailed in section \ref{section : A systematic multi-tone jamming approach} provides an efficient solution against the Wiener interpolation filter, under perfect knowledge or real estimates of the correlation functions. Last but not least, these results illustrate how the filter length $L$ is a sensitive information for the receiver security since the number of tones $K$ can be adapted to make the anti-jamming module completely ineffective.

\section{Conclusion}
\label{section : Conclusion}
In this paper, a method for generating the most difficult-to-estimate $K$-tone jammer for a $L$-tap Wiener interpolation filter is proposed. The core principle of this method is to maximize the BMSE of the jamming waveform estimate produced by the Wiener interpolation filter. To this end, closed-form expressions of both the Wiener interpolation filter and the BMSE of the multi-tone estimate are derived analytically. Based on these expressions, the optimal two-tone jammer is obtained in closed form. This jamming waveform, characterized by two equal-power tones with a normalized angular frequency separation $\Delta_{21}^{\dagger}$ of $9/(L+1)$ or $2\pi-9/(L+1)$, jointly maximizes the BMSE of the individual tone estimates and the correlation of estimation errors. This analytical result is then generalized to the $K$-tone case and refined by introducing an additional frequency correction term per tone, which further optimizes the correlation of estimation errors between non-adjacent tones. Additionally, this paper shows that a jammer composed of equal-power tones, whose normalized angular frequencies are equally spaced by $2\pi/K$, becomes optimal when $K\geq L/2+1$ tones. Overall, this work highlights that the filter length $L$ is a critical parameter from a security perspective. If this parameter is known, low-complexity narrow-band devices can be employed to render the Wiener interpolation filter used in wideband systems completely ineffective.

\appendix
\numberwithin{equation}{section}
\setcounter{equation}{0}
\subsection{Derivation of the analytical form of the Wiener interpolation filter $\mathbf{w}^{\star}$ for a multi-tone jammer in DSSS systems}
\label{appendix : Wiener Filter}
 From \eqref{eqn : i_m est selon w}, \eqref{eqn : i_m est selon theta} and \eqref{eqn : theta_m est}, the general form of the Wiener filter is $\mathbf{w}^{\star}=\mathbf{C}_{\mathbf{r}\mathbf{r}}^{-1}(\boldsymbol{\alpha},\boldsymbol{\omega})\mathbf{C}_{\mathbf{r}\boldsymbol{\theta}}(\boldsymbol{\alpha},\boldsymbol{\omega})\boldsymbol{1}_K$. To obtain the special form of the Wiener interpolation filter for a multi-tone jammer in DSSS systems, the matrices $\mathbf{C}_{\mathbf{rr}}(\boldsymbol{\alpha},\boldsymbol{\omega})$ and $\mathbf{C}_{\mathbf{r}\boldsymbol{\theta}}(\boldsymbol{\alpha},\boldsymbol{\omega})$ are respectively derived in Appendices \ref{appendix : Crr calculation} and \ref{appendix : Crtheta calculation}.


\subsubsection{Derivation of the matrix $\mathbf{C}_{\mathbf{r}\mathbf{r}}(\boldsymbol{\alpha},\boldsymbol{\omega})$ }
\label{appendix : Crr calculation}
Firstly, it is noticed that $\mathbf{r}_m(\boldsymbol{\alpha},\boldsymbol{\omega})\in\mathbb{C}^K$ is composed of random variables $\mathbf{s}_m\in\mathbb{C}^K$, $\mathbf{n}_m\in\mathbb{C}^K$ and $\mathbf{i}_m(\boldsymbol{\alpha},\boldsymbol{\omega})\in\mathbb{C}^K$ that are mutually uncorrelated. Additionally, $\mathbf{s}_m$ and $\mathbf{n}_m$ are independent and identically distributed. Consequently,
\begin{align}
\label{eqn : C_rr calculation}
    &\mathbf{C}_{\mathbf{r}\mathbf{r}}(\boldsymbol{\alpha},\boldsymbol{\omega})=\mathbb{E}_{s,n,\phi}\left[\mathbf{r}_m(\boldsymbol{\alpha},\boldsymbol{\omega})\mathbf{r}_m^H(\boldsymbol{\alpha},\boldsymbol{\omega})\right],\nonumber\\
    &=\mathbb{E}_s\left[\mathbf{s}_m\mathbf{s}_m^H\right]+\mathbb{E}_n\left[\mathbf{n}_m\mathbf{n}_m^H\right]+\mathbb{E}_{\phi}\left[\mathbf{i}_m(\boldsymbol{\alpha},\boldsymbol{\omega})\mathbf{i}_m^H(\boldsymbol{\alpha},\boldsymbol{\omega})\right],\nonumber\\
    &=(S+\sigma_n^2)\times\mathbf{I}_L+\mathbb{E}_{\phi}\left[\mathbf{i}_m(\boldsymbol{\alpha},\boldsymbol{\omega})\mathbf{i}_m^H(\boldsymbol{\alpha},\boldsymbol{\omega})\right].
\end{align}
From \eqref{eq :i_m} and \eqref{eqn : theta_m}, it is remarked that $\mathbf{i}_m(\boldsymbol{\alpha},\boldsymbol{\omega})$ can be written, 
\begin{equation}
    \label{eqn : i_m vector form}\mathbf{i}_m(\boldsymbol{\alpha},\boldsymbol{\omega})=\boldsymbol{\Psi}_m(\boldsymbol{{\omega}})\boldsymbol{\beta}(\boldsymbol{\alpha}),
\end{equation}
where,
\begin{equation}
\boldsymbol{\Psi}_m(\boldsymbol{{\omega}})=\left[\boldsymbol{\psi}_m({\omega_1}),\boldsymbol{\psi}_m({\omega_2}),\ldots,\boldsymbol{\psi}_m({\omega_K})]\right]\in\mathbb{C}^{L\times K},
\end{equation}
\begin{align}
\boldsymbol{\psi}_m({\omega_k})&=\left[e^{j\left(m+\frac{L}{2}\right)\omega_k},\ldots,e^{j(m+1)\omega_k},\right.\nonumber\\&\qquad\left.e^{j(m-1)\omega_k},\ldots,e^{j\left(m-\frac{L}{2}\right)\omega_k}\right]^T\in\mathbb{C}^{L},
\end{align}
\begin{equation}
\label{eqn : beta vector form}
    \boldsymbol{\beta}(\boldsymbol{\alpha})=\left[\alpha_1 e^{j\phi_1},\alpha_2 e^{j\phi_2},\ldots,\alpha_K e^{j\phi_K}\right]^T\in\mathbb{C}^K.
\end{equation}
Therefore, the term $\mathbb{E}_{\phi}\left[\mathbf{i}_m\mathbf{i}_m^H\right]$ can be expressed,
\begin{align}
\label{eqn : C_ri calculation}
\mathbb{E}_{\phi}\left[\mathbf{i}_m\mathbf{i}_m^H\right]&=\boldsymbol{\Psi}_m(\boldsymbol{{\omega}})\mathbb{E}_{\phi}\left[\boldsymbol{\beta}(\boldsymbol{\alpha})\boldsymbol{\beta}^H(\boldsymbol{\alpha})\right]\boldsymbol{\Psi}_m^H(\boldsymbol{{\omega}}),\nonumber\\
    &=\boldsymbol{\Psi}_m(\boldsymbol{{\omega}})\mathbf{J}(\boldsymbol{\alpha})\boldsymbol{\Psi}_m^H(\boldsymbol{{\omega}}),\nonumber\\
&=\boldsymbol{\Psi}(\boldsymbol{{\omega}})\mathbf{J}(\boldsymbol{\alpha})\boldsymbol{\Psi}^H(\boldsymbol{{\omega}})
\end{align}
where $\mathbf{J}(\boldsymbol{\alpha})$ and $\boldsymbol{\Psi}(\boldsymbol{{\omega}})$ are given by \eqref{eqn : J_matrix}-\eqref{eqn : psi_k}. Finally, the analytical expression of $\mathbf{C}_{\mathbf{r}\mathbf{r}}(\boldsymbol{\alpha},\boldsymbol{\omega})$ is obtained by substituting \eqref{eqn : C_ri calculation} into \eqref{eqn : C_rr calculation}, yielding \eqref{eqn : C_rr}. 


\subsubsection{Derivation of the matrix $\mathbf{C}_{\mathbf{r}\boldsymbol{\theta}}(\boldsymbol{\alpha},\boldsymbol{\omega})$}
\label{appendix : Crtheta calculation}
From \eqref{eqn : theta_m}, \eqref{eqn : i_m vector form} and the mutual uncorrelatedness of $\mathbf{s}_m$, $\mathbf{n}_m$ and $\mathbf{i}_m(\boldsymbol{\alpha},\boldsymbol{\omega})$,
\begin{align}
    \mathbf{C}_{\mathbf{r}\boldsymbol{\theta}}(\boldsymbol{\alpha},\boldsymbol{\omega}) &= \mathbb{E}_{s,n,\phi}\left[\mathbf{r}_m(\boldsymbol{\alpha},\boldsymbol{\omega})\boldsymbol{\theta}_m^H(\boldsymbol{\alpha},\boldsymbol{\omega})\right],\nonumber\\
    &= \mathbb{E}_{\phi}\left[\mathbf{i}_m(\boldsymbol{\alpha},\boldsymbol{\omega})\boldsymbol{\theta}_m^H(\boldsymbol{\alpha},\boldsymbol{\omega})\right],\nonumber\\
    &= \mathbb{E}_{\phi}\left[\boldsymbol{\Psi}_m(\boldsymbol{{\omega}})\boldsymbol{\beta}(\boldsymbol{\alpha})\boldsymbol{\theta}_m^H(\boldsymbol{\alpha},\boldsymbol{\omega})\right],\nonumber\\
    &=\boldsymbol{\Psi}_m(\boldsymbol{{\omega}})\mathrm{diag}\left(\alpha_1^2e^{-j\omega_1m},\ldots,\alpha_K^2e^{-j\omega_Km}\right),\nonumber\\
    &=\boldsymbol{\Psi}(\boldsymbol{{\omega}})\mathbf{J}(\boldsymbol{\alpha}),
\end{align}
which concludes the analytical derivation of $\mathbf{C}_{\mathbf{r}\boldsymbol{\theta}}(\boldsymbol{\alpha},\boldsymbol{\omega})$ \eqref{eqn : C_rtheta} and the proof of Theorem \ref{th : wiener filter}. 

\subsection{Derivation of the analytical BMSE of the interference estimate $\hat{i}^{\star}(\boldsymbol{\alpha},\boldsymbol{\omega})$ for a multi-tone jammer in DSSS systems}
\label{eqn : bayesian MSE}
\renewcommand{\theequation}{B.\arabic{equation}}
\setcounter{equation}{0}
From the general expression of the BMSE \eqref{eqn : BMSE i est} and the expressions of ${i}_m(\boldsymbol{\alpha},\boldsymbol{\omega})$ and $\hat{i}_m^{\star}(\boldsymbol{\alpha},\boldsymbol{\omega})$, respectively given in \eqref{eq :i_m} and \eqref{eqn : i_m est selon theta}, the BMSE of the interference estimate $\hat{i}^{\star}(\boldsymbol{\alpha},\boldsymbol{\omega})$ can be expressed as,
\begin{align}
\label{eqn : appendix bmse i star}
    \mathrm{BMSE}\big(\,\hat{i}^{\star}(\boldsymbol{\alpha},\boldsymbol{\omega})\,\big) &= \mathbb{E}_{s,n,\phi}\left[|i_m(\boldsymbol{\alpha},\boldsymbol{\omega})-\hat{i}_m^{\star}(\boldsymbol{\alpha},\boldsymbol{\omega})|^2\right],\nonumber\\
    &=\mathbb{E}_{s,n,\phi}\left[\Big|\boldsymbol{1}_K^T\boldsymbol{\epsilon}(\boldsymbol{\alpha},\boldsymbol{\omega})\Big|^2\right],\nonumber\\
    &=\boldsymbol{1}_K^T\mathbf{C}_{{\boldsymbol{\epsilon}}}^{\star}(\boldsymbol{\alpha},\boldsymbol{\omega})\boldsymbol{1}_K, 
\end{align}
where $\boldsymbol{\epsilon}(\boldsymbol{\alpha},\boldsymbol{\omega})=\boldsymbol{\theta}_m(\boldsymbol{\alpha},\boldsymbol{\omega})-\hat{\boldsymbol{\theta}}_m^{\star}(\boldsymbol{\alpha},\boldsymbol{\omega})$ denotes the estimation error between each tone contribution $\boldsymbol{\theta}_m(\boldsymbol{\alpha},\boldsymbol{\omega})$ and its estimate $\hat{\boldsymbol{\theta}}_m^{\star}(\boldsymbol{\alpha},\boldsymbol{\omega})$, while $\mathbf{C}_{\boldsymbol{\epsilon}}^{\star}(\boldsymbol{\alpha},\boldsymbol{\omega})$ characterizes the minimum error covariance matrix. The latter can be written \cite[eq.(12.21)]{Funda_Stat_Signal_Processing_EstTheory},
\begin{align}
    \label{eqn : appendix C_esp star general form}\mathbf{C}_{\boldsymbol{\epsilon}}^{\star}(\boldsymbol{\alpha},\boldsymbol{\omega})&=\mathbb{E}_{s,n,\phi}\left[\boldsymbol{\epsilon}(\boldsymbol{\alpha},\boldsymbol{\omega})\boldsymbol{\epsilon}^H(\boldsymbol{\alpha},\boldsymbol{\omega})\right],\nonumber\\
    &=\mathbf{C}_{\boldsymbol{\theta}\boldsymbol{\theta}}(\boldsymbol{\alpha},\boldsymbol{\omega})-\mathbf{C}_{\mathbf{r}\boldsymbol{\theta}}^H(\boldsymbol{\alpha},\boldsymbol{\omega})\mathbf{C}_{\mathbf{r}\mathbf{r}}^{-1}(\boldsymbol{\alpha},\boldsymbol{\omega})\mathbf{C}_{\mathbf{r}\boldsymbol{\theta}}(\boldsymbol{\alpha},\boldsymbol{\omega}),\nonumber\\
\end{align}
where the covariance matrix $\mathbf{C}_{\boldsymbol{\theta}\boldsymbol{\theta}}(\boldsymbol{\alpha},\boldsymbol{\omega})$ reduces to,
\begin{align}
    \label{eqn : appendix C theta}\mathbf{C}_{\boldsymbol{\theta}\boldsymbol{\theta}}(\boldsymbol{\alpha},\boldsymbol{\omega})&=\mathbb{E}_{\phi}\left[\boldsymbol{\theta}_m(\boldsymbol{\alpha},\boldsymbol{\omega})\boldsymbol{\theta}_m^H(\boldsymbol{\alpha},\boldsymbol{\omega})\right],\nonumber\\
    &=\mathbf{J}(\boldsymbol{\alpha}).
\end{align}
Consequently, from \eqref{eqn : appendix C_esp star general form}, \eqref{eqn : appendix C theta}, \eqref{eqn : C_rr} and \eqref{eqn : C_rtheta},
\begin{align}
    &\mathbf{C}_{\boldsymbol{\epsilon}}^{\star}(\boldsymbol{\alpha},\boldsymbol{\omega})=\mathbf{J}(\boldsymbol{\alpha})-\mathbf{J}(\boldsymbol{\alpha})\boldsymbol{\Psi}^H(\boldsymbol{\omega})\times\nonumber\\&\left((S+\sigma_n^2)\mathbf{I}_L+\boldsymbol{\Psi}(\boldsymbol{\omega})\mathbf{J}(\boldsymbol{\alpha})\boldsymbol{\Psi}^H(\boldsymbol{\omega})\right)^{-1}\boldsymbol{\Psi}(\boldsymbol{\omega})\mathbf{J}(\boldsymbol{\alpha}).
\end{align}
To further simplify this equation, the matrix $\mathbf{B}$ is first defined by setting $\mathbf{B}=\mathbf{J}^{-1}(\boldsymbol{\alpha})+(S+\sigma_n^2)^{-1}\boldsymbol{\Gamma}(\boldsymbol{\omega})$, where $\boldsymbol{\Gamma}(\boldsymbol{\omega})=\boldsymbol{\Psi}^H(\boldsymbol{\omega})\boldsymbol{\Psi}(\boldsymbol{\omega})$. It's worth noting that $\left[\boldsymbol{\Gamma}(\boldsymbol{\omega})\right]_{kk'}$ allows a closed-form expression given by, 
\begin{align}
    \left[\boldsymbol{\Gamma}(\boldsymbol{\omega})\right]_{kk'}&=\boldsymbol{\psi}^H({\omega_k})\boldsymbol{\psi}({\omega_{k'}}),\nonumber\\
    &=\sum_{\underset{l\neq 0}{l=-L/2}}^{L/2}{e^{jl({\omega}_{k'}-{\omega}_k)}},\nonumber\\
    &=\frac{\sin{\left(\frac{L+1}{2}\Delta_{k'k}\right)}}{\sin{\left(\frac{1}{2}\Delta_{k'k}\right)}}-1,\nonumber\\
    &=D_{L/2}\left(\Delta_{k'k}\right)-1,
\end{align}
where $\Delta_{k'k}=w_{k'}-w_k$ is the normalized angular frequency difference between the $k'$th tone and the $k$th one, and $D_{n}(x)=\sin{\left((n+1/2)x\right)}/\sin{\left(x/2\right)}$ denotes the $n$th Dirichlet kernel.
After using the Woodbury identity, the matrix $\mathbf{C}_{\boldsymbol{\epsilon}}^{\star}(\boldsymbol{\alpha},\boldsymbol{\omega})$ reduces to,
\begin{align}
\label{eqn: appendix C epsilon}
    &\mathbf{C}_{\boldsymbol{\epsilon}}^{\star}(\boldsymbol{\alpha},\boldsymbol{\omega})=\mathbf{J}(\boldsymbol{\alpha})-\mathbf{J}(\boldsymbol{\alpha})\boldsymbol{\Psi}^H(\boldsymbol{\omega})\times\nonumber\\ &\left[(S+\sigma_n^2)^{-1}\mathbf{I}_L-(S+\sigma_n^2)^{-2}\boldsymbol{\Psi}(\boldsymbol{\omega})\mathbf{B}^{-1}\boldsymbol{\Psi}^H(\boldsymbol{\omega})\right]\boldsymbol{\Psi}(\boldsymbol{\omega})\mathbf{J}(\boldsymbol{\alpha}),\nonumber\\
    &\qquad\qquad=\mathbf{J}(\boldsymbol{\alpha})-\mathbf{J}(\boldsymbol{\alpha})\left[\left(\mathbf{B}-\mathbf{J}^{-1}(\boldsymbol{\alpha})\right)-\right.\nonumber\\&\quad\left.\left(\mathbf{B}-\mathbf{J}^{-1}(\boldsymbol{\alpha})\right)\mathbf{B}^{-1}\left(\mathbf{B}-\mathbf{J}^{-1}(\boldsymbol{\alpha})\right)\right]\mathbf{J}(\boldsymbol{\alpha}),\nonumber\\
    &\qquad\qquad=\mathbf{B}^{-1}.
\end{align}
Finally, \eqref{eqn : C_espsilon star} and \eqref{eqn : A matrix} are obtained by substituting \eqref{eqn: appendix C epsilon} into \eqref{eqn : appendix bmse i star}, which concludes the proof of Theorem \ref{th : analytical BMSE}. 

\subsection{Closed-form derivation of the partial derivatives of $\mathcal{L}(\boldsymbol{\alpha},\boldsymbol{\omega},\lambda)$}
\label{Calculation of partial derivatives}
\renewcommand{\theequation}{C.\arabic{equation}}
This section details the closed-form derivation of ${\partial \mathcal{L}(\boldsymbol{\alpha},\boldsymbol{\omega},\lambda)}/{\partial \alpha_k^2}$ in Appendix \ref{appendix : partial derivative of the loss function with regard to alpha_i^2} and ${\partial \mathcal{L}(\boldsymbol{\alpha},\boldsymbol{\omega},\lambda)}/{\partial \omega_k}$ in Appendix \ref{appendix : partial derivative of the loss function with regard to omega_i}. 
\setcounter{equation}{0}

\subsubsection{Derivation of ${\partial \mathcal{L}(\boldsymbol{\alpha},\boldsymbol{\omega},\lambda)}/{\partial \alpha_k^2}$}
\label{appendix : partial derivative of the loss function with regard to alpha_i^2}
From the differential properties $\mathrm{d}(\mathbf{XY})=\mathrm{d}(\mathbf{X})\mathbf{Y}+\mathbf{X}\mathrm{d}(\mathbf{Y})$ and $\mathrm{d}\mathbf{X}^{-1}=-\mathbf{X}^{-1}(\mathrm{d}\mathbf{X})\mathbf{X}^{-1}$, the partial derivatives of $\mathbf{XY}$ and $\mathbf{X}^{-1}$ with respect to $\alpha_k^2$ are respectively written \cite{Minka2000OldAN},    
\begin{equation}
\label{eqn : partial derivative of a matrix product}
    \frac{\partial(\mathbf{XY})}{\partial \alpha_k^2}=\frac{\partial\mathbf{X}}{\partial \alpha_k^2}\mathbf{Y}+\mathbf{X}\frac{\partial\mathbf{Y}}{\partial \alpha_k^2},
\end{equation}
\begin{equation}
\label{eqn : partial derivative of matrix inverse}
    \frac{\partial(\mathbf{X}^{-1})}{\partial \alpha_k^2}=-\mathbf{X}^{-1}\frac{\partial\mathbf{X}}{\partial \alpha_k^2}\mathbf{X}^{-1}.
\end{equation}
Therefore, from \eqref{eqn : BMSE i est star}, \eqref{eqn : C_espsilon star} and \eqref{eqn : partial derivative of a matrix product}, the partial derivative of $\mathcal{L}(\boldsymbol{\alpha},\boldsymbol{\omega},\lambda)$ with respect to $\alpha_k^2$ is written,
\begin{align}
\label{eqn : appendix partial derivative of BMSE according to alpha_i^2 1st eq}
     \frac{\partial \mathcal{L}(\boldsymbol{\alpha},\boldsymbol{\omega},\lambda)}{\partial \alpha_k^2}&=-\boldsymbol{1}_K^T\left(\frac{\partial \mathbf{A}^{-1}(\boldsymbol{\alpha},\boldsymbol{\omega})}{\partial \alpha_k^2}\mathbf{J}(\boldsymbol{\alpha})+\right.\nonumber\\&\qquad\qquad\left.\mathbf{A}^{-1}(\boldsymbol{\alpha},\boldsymbol{\omega})\frac{\partial \mathbf{J}(\boldsymbol{\alpha})}{\partial \alpha_k^2}\right)\boldsymbol{1}_K+\lambda,
\end{align}
where,
\begin{equation}
\label{eqn : partial derivative J with respect to alpha_k}
    \frac{\partial \mathbf{J}(\boldsymbol{\alpha})}{\partial \alpha_k^2}=\mathrm{diag}(0,\ldots,0,\underbrace{1}_{\substack{k^{\mathrm{th}}\\ \mathrm{element}}},0,\ldots,0). 
\end{equation}

Additionally, using property \eqref{eqn : partial derivative of matrix inverse}, the term ${\partial \mathbf{A}^{-1}(\boldsymbol{\alpha},\boldsymbol{\omega})}/{\partial \alpha_k^2}$ is expressed as,
\begin{align}
\label{eqn : partial derivative of inv(A) according to alpha_i^2}
    &\frac{\partial \mathbf{A}^{-1}(\boldsymbol{\alpha},\boldsymbol{\omega})}{\partial \alpha_k^2}=-\mathbf{A}^{-1}(\boldsymbol{\alpha},\boldsymbol{\omega})\frac{\partial\mathbf{A}(\boldsymbol{\alpha},\boldsymbol{\omega})}{\partial \alpha_k^2}\mathbf{A}^{-1}(\boldsymbol{\alpha},\boldsymbol{\omega}),\nonumber\\
    &=-(S+\sigma_n^2)^{-1}\mathbf{A}^{-1}(\boldsymbol{\alpha},\boldsymbol{\omega})\frac{\partial \mathbf{J}(\boldsymbol{\alpha})}{\partial \alpha_k^2}\boldsymbol{\Gamma}(\boldsymbol{\omega})\mathbf{A}^{-1}(\boldsymbol{\alpha},\boldsymbol{\omega}).
\end{align}
Then, by substituting \eqref{eqn : partial derivative of inv(A) according to alpha_i^2} into \eqref{eqn : appendix partial derivative of BMSE according to alpha_i^2 1st eq} and exploiting the matrix inverse property given in \cite[eq.(166)]{Petersen2006TheMC},
\begin{align}
\label{eqn : appendix partial derivative of BMSE according to alpha_i^2 2nd eq}
    &\frac{\partial \mathcal{L}(\boldsymbol{\alpha},\boldsymbol{\omega},\lambda)}{\partial \alpha_k^2}=-\boldsymbol{1}_K^T\mathbf{A}^{-1}(\boldsymbol{\alpha,\boldsymbol{\omega}})\frac{\partial \mathbf{J}(\boldsymbol{\alpha})}{\partial \alpha_k^2}\times\\&\left(\mathbf{I}_K-\frac{\boldsymbol{\Gamma}(\boldsymbol{\omega})}{S+\sigma_n^2} \left(\mathbf{I}_K+\mathbf{J}(\boldsymbol{\alpha})\frac{\boldsymbol{\Gamma}(\boldsymbol{\omega})}{S+\sigma_n^2}\right)^{-1}\mathbf{J}(\boldsymbol{\alpha})\right)\boldsymbol{1}_K+\lambda,\nonumber\\
    &=-\boldsymbol{1}_K^T\mathbf{A}^{-1}(\boldsymbol{\alpha,\boldsymbol{\omega}})\frac{\partial \mathbf{J}(\boldsymbol{\alpha})}{\partial \alpha_k^2}\left(\mathbf{I}_K+\frac{\boldsymbol{\Gamma}(\boldsymbol{\omega})}{S+\sigma_n^2}\mathbf{J}(\boldsymbol{\alpha})\right)^{-1}\boldsymbol{1}_K+\lambda,\nonumber\\
    &= -\boldsymbol{1}_K^T\mathbf{A}^{-1}(\boldsymbol{\alpha,\boldsymbol{\omega}})\frac{\partial \mathbf{J}(\boldsymbol{\alpha})}{\partial \alpha_k^2}\mathbf{A}^{-T}(\boldsymbol{\alpha,\boldsymbol{\omega}})\boldsymbol{1}_K+\lambda,
\end{align}
which results in \eqref{eqn : partial derivative of BMSE according to alpha_i^2} and \eqref{eqn : E matrix} by substituting \eqref{eqn : partial derivative J with respect to alpha_k} into \eqref{eqn : appendix partial derivative of BMSE according to alpha_i^2 2nd eq}. 

\subsubsection{Derivation of ${\partial \mathcal{L}(\boldsymbol{\alpha},\boldsymbol{\omega},\lambda)}/{\partial \omega_k}$}
\label{appendix : partial derivative of the loss function with regard to omega_i}
From \eqref{eqn : BMSE i est star}, \eqref{eqn : C_espsilon star} and \eqref{eqn : partial derivative of a matrix product}, the partial derivative of $\mathcal{L}(\boldsymbol{\alpha},\boldsymbol{\omega},\lambda)$ with respect to $\omega_k$ is written,
\begin{align}
\label{appendix : partial derivative of loss function with respect to omega_i}
    \frac{\partial \mathcal{L}(\boldsymbol{\alpha},\boldsymbol{\omega},\lambda)}{\partial \omega_k}&=-\boldsymbol{1}_K^T\frac{\partial(\mathbf{A}^{-1}(\boldsymbol{\alpha},\boldsymbol{\omega})\mathbf{J}(\boldsymbol{\alpha}))}{\partial \omega_k}\boldsymbol{1}_K,\nonumber\\
    &=-\boldsymbol{1}_K^T\frac{\partial \mathbf{A}^{-1}(\boldsymbol{\alpha},\boldsymbol{\omega})}{\partial \omega_k}\mathbf{J}(\boldsymbol{\alpha})\boldsymbol{1}_K.
\end{align}
Using \eqref{eqn : partial derivative of matrix inverse}, ${\partial \mathbf{A}^{-1}(\boldsymbol{\alpha},\boldsymbol{\omega})}/{\partial \omega_k}$ can be written as,
\begin{equation}
\label{eqn : partial derivative of inv A with respect to omega_i}
\frac{\partial \mathbf{A}^{-1}(\boldsymbol{\alpha},\boldsymbol{\omega})}{\partial \omega_k}=-\frac{\mathbf{A}^{-1}(\boldsymbol{\alpha},\boldsymbol{\omega})}{S+\sigma_n^2}\mathbf{J}(\boldsymbol{\alpha})\frac{\partial\boldsymbol{\Gamma}(\boldsymbol{\omega})}{\partial \omega_k}\mathbf{A}^{-1}(\boldsymbol{\alpha},\boldsymbol{\omega}),
\end{equation}
where ${\boldsymbol{\partial\Gamma}(\boldsymbol{\omega})}/{\partial \omega_k}$ is expressed as, 
\begin{align}
    \frac{\partial\boldsymbol{\Gamma}(\boldsymbol{\omega})}{\partial \omega_k} &= \frac{\partial}{\partial \omega_k}
    \begin{bmatrix}
L & \boldsymbol{\Gamma}_{12} & \ldots & \boldsymbol{\Gamma}_{1k} & \ldots & \boldsymbol{\Gamma}_{1K}\\
\boldsymbol{\Gamma}_{21} & L & \ldots & \boldsymbol{\Gamma}_{2k} & \ldots & \boldsymbol{\Gamma}_{2K}\\
 \vdots& \vdots & \ddots & \vdots & \ldots & \vdots\\
 \boldsymbol{\Gamma}_{k1} & \boldsymbol{\Gamma}_{k2} & \ldots & L & \ldots & \boldsymbol{\Gamma}_{kK}\\
 \vdots& \vdots &  & \vdots & \ddots & \vdots\\
  \boldsymbol{\Gamma}_{K1}& \boldsymbol{\Gamma}_{K2} & \ldots & \boldsymbol{\Gamma}_{Kk}  & \ldots & L\\
\end{bmatrix},\nonumber\\
&=\quad\
    \begin{bmatrix}
0 & 0 & \ldots & \frac{\partial\boldsymbol{\Gamma}_{1k}}{\partial\omega_k} & \ldots & 0\\
0 & 0 & \ldots & \frac{\partial\boldsymbol{\Gamma}_{2k}}{\partial\omega_k}& \ldots & 0\\
 \vdots& \vdots & \ddots & \vdots & \ldots & \vdots\\
 \frac{\partial\boldsymbol{\Gamma}_{k1}}{\partial\omega_k} & \frac{\partial\boldsymbol{\Gamma}_{k2}}{\partial\omega_k} & \ldots & 0 & \ldots & \frac{\partial\boldsymbol{\Gamma}_{kK}}{\partial\omega_k}\\
 \vdots& \vdots &  & \vdots & \ddots & \vdots\\
  0& 0 & \ldots & \frac{\partial\boldsymbol{\Gamma}_{Kk}}{\partial\omega_k}  & \ldots & 0\\
\end{bmatrix}. 
\end{align}
In this matrix, elements $\left[\frac{\partial \boldsymbol{\Gamma}(\boldsymbol{\omega})}{\partial \omega_k}\right]_{kk^{'}}$ and $\left[\frac{\partial \boldsymbol{\Gamma}(\boldsymbol{\omega})}{\partial \omega_k}\right]_{k^{'}k}$ are derived from \eqref{eqn : Gamma_ij} to give \eqref{eqn : partial derivative of gamma matrix with respect to omega_i} and \eqref{eqn : partial derivative of the Dirichlet kernel}. Finally, \eqref{eqn : partial derivative of inv A with respect to omega_i} is substituted into \eqref{appendix : partial derivative of loss function with respect to omega_i} to obtain \eqref{eqn : partial derivative of the loss function with respect to omega_i}, ending the proof of Theorem \ref{th : partial derivatives}.   
\subsection{Optimization of a 2-tone jammer}
\label{2-tone jamming signal optimization}
In this section, the closed-form expression of the optimal 2-tone jammer is derived. To this end, the optimal solution $\boldsymbol{\alpha}^{\dagger}$ is first established in Appendix \ref{appendix : proof alpha dagger}, before deriving the optimal solution $\boldsymbol{\omega}^{\dagger}$ in Appendix \ref{appendix : proof omega dagger}. 
\renewcommand{\theequation}{D.\arabic{equation}}
\setcounter{equation}{0}
\subsubsection{Proof that $\boldsymbol{\alpha}^{\dagger}=\left[\sqrt{J/2},\sqrt{J/2}\right]^T$}
\label{appendix : proof alpha dagger}
First, the convexity of the Lagrangian function $\mathcal{L}(\boldsymbol{\alpha},\boldsymbol{\omega},\lambda)$ with respect to  $\alpha_k^2>0$ is examined, treating $\boldsymbol{\omega}$ as a fixed parameter. From \eqref{eqn : A matrix}, \eqref{eqn : E matrix} and the formula of the inverse for a $2\times 2$ matrix, coefficients $e_{kk^{'}}=\left[\mathbf{E}(\boldsymbol{\alpha},\boldsymbol{\omega})\right]_{kk^{'}}$ are given by,
\begin{equation}
\label{eqn : E elements}
    \left\{ 
    \begin{array}{l}
            e_{11}=\nu(\alpha_1^2,\alpha_2^2)\times\left(1+\alpha_2^2L/(S+\sigma_n^2)\right) \\
            e_{22}=\nu(\alpha_1^2,\alpha_2^2)\times\left(1+\alpha_1^2L/(S+\sigma_n^2)\right)\\
            e_{21}=-\nu(\alpha_1^2,\alpha_2^2)\times\alpha_2^2\left(D_{L/2}(\Delta_{21})-1\right)/(S+\sigma_n^2)\\
            e_{12}=-\nu(\alpha_1^2,\alpha_2^2)\times\alpha_1^2\left(D_{L/2}(\Delta_{21})-1\right)/(S+\sigma_n^2)\\
            
            \nu(\alpha_1^2,\alpha_2^2) = \left[
            1+(\alpha_1^2+\alpha_2^2)L/(S+\sigma_n^2)+\alpha_1^2\alpha_2^2\times\right.\\\qquad\qquad\left.\left[L^2-(D_{L/2}(\Delta_{21})-1)^2\right]/(S+\sigma_n^2)^2\right]^{-1}
    \end{array}
    \right. .
\end{equation}
Partial derivatives of $\mathcal{L}(\boldsymbol{\alpha},\boldsymbol{\omega},\lambda)$ with respect to $\alpha_k^2$ are then calculated from \eqref{eqn : partial derivative of BMSE according to alpha_i^2} and become,
\begin{equation}
\label{eqn : appendix partial derivative according to alpha_1^2}
\left\{ 
    \begin{array}{l}
    \frac{\partial \mathcal{L}(\boldsymbol{\alpha},\boldsymbol{\omega},\lambda)}{\partial \alpha_1^2} =-\left[\nu(\alpha_1^2,\alpha_2^2)\right]^2\times\left[1+\alpha_2^2\frac{\gamma}{S+\sigma_n^2}\right]^2+\lambda\\
    \frac{\partial \mathcal{L}(\boldsymbol{\alpha},\boldsymbol{\omega},\lambda)}{\partial \alpha_2^2} =-\left[\nu(\alpha_1^2,\alpha_2^2)\right]^2\times\left[1+\alpha_1^2\frac{\gamma}{S+\sigma_n^2}\right]^2+\lambda
    \end{array}
    \right. ,
\end{equation}
where $\gamma=L-\left(D_{L/2}(\Delta_{21})-1\right)\geq0$. Recalling that the Lagrangian function $\mathcal{L}(\boldsymbol{\alpha},\boldsymbol{\omega},\lambda)$ is convex with respect to $\alpha_k^2$ if and only if ${\partial \mathcal{L}(\boldsymbol{\alpha},\boldsymbol{\omega},\lambda)}/{\partial \alpha_k^2}$ is an increasing function, the monotonicity of \eqref{eqn : appendix partial derivative according to alpha_1^2} is studied. From the analytical expression of $\nu(\alpha_1^2,\alpha_2^2)$ \eqref{eqn : E elements}, the function $-\left[\nu(\alpha_1^2,\alpha_2^2)\right]^2$ is an increasing function for $\alpha_1^2>0$ (resp. $\alpha_2^2>0$). Therefore, ${\partial \mathcal{L}(\boldsymbol{\alpha},\boldsymbol{\omega},\lambda)}/{\partial \alpha_1^2}$ (resp. ${\partial \mathcal{L}(\boldsymbol{\alpha},\boldsymbol{\omega},\lambda)}/{\partial \alpha_2^2}$) is also an increasing function over the same interval. As a consequence, $\mathcal{L}(\boldsymbol{\alpha},\boldsymbol{\omega},\lambda)$ is convex on $\alpha_1^2>0$ and $\alpha_2^2>0$. The Lagrangian function is thus minimized when ${\partial \mathcal{L}(\boldsymbol{\alpha},\boldsymbol{\omega},\lambda)}/{\partial \alpha_1^2}={\partial \mathcal{L}(\boldsymbol{\alpha},\boldsymbol{\omega},\lambda)}/{\partial \alpha_2^2}=0$ and ${\partial \mathcal{L}(\boldsymbol{\alpha},\boldsymbol{\omega},\lambda)}/{\partial \lambda}=0$, which are obtained for $\alpha_1^2=\alpha_2^2=J/2$, leading to the final result \eqref{eqn : alpha dagger}. 

\subsubsection{Proof that $\boldsymbol{\omega}^{\dagger}=\left[\omega_1,\omega_1+\Delta_{21}^{\dagger}\right]^T$}
\label{appendix : proof omega dagger}
\begin{figure}[!t]
    \centering
    \subfloat[$\Delta_{21}^{\dagger(1)}$ approximate solution.]{%
        \includegraphics[width=0.5\textwidth]{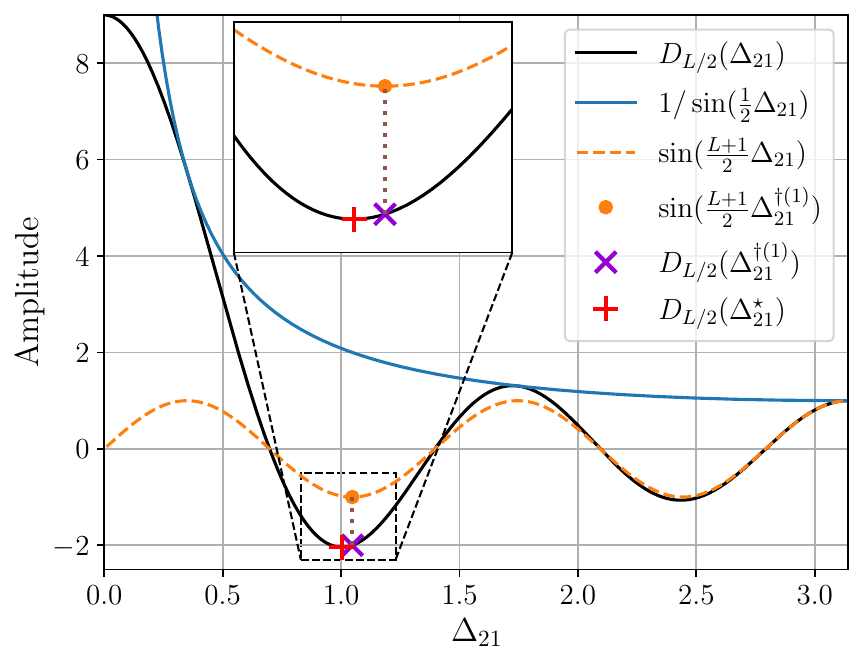}%
        \label{fig: Delta_21 approx 1}
    }\hfill
    \subfloat[Absolute difference $\epsilon_{\Delta_{21}^{\star}}$ with respect to the filter length $L$]{%
        \includegraphics[width=0.5\textwidth]{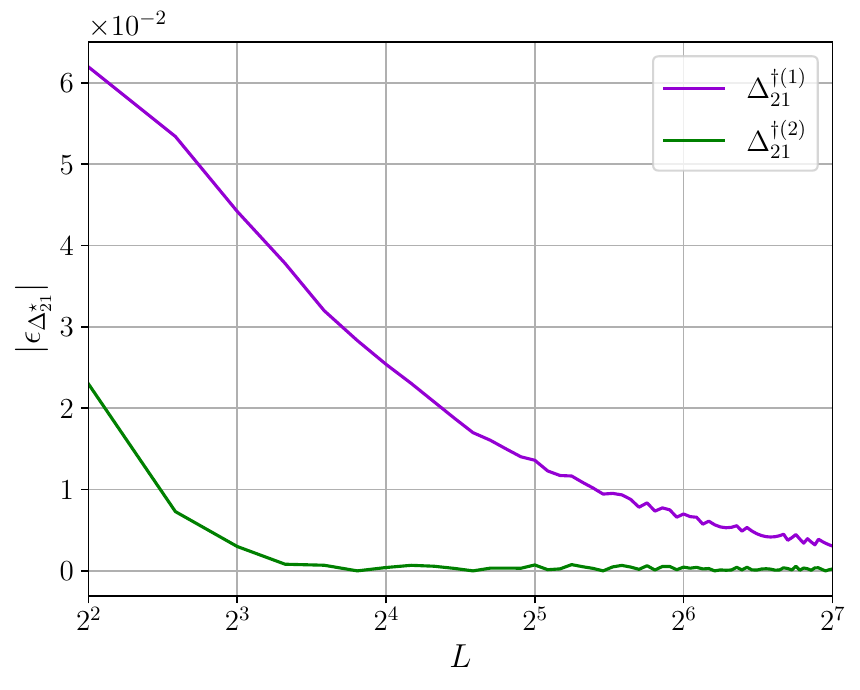}%
        \label{fig:subfig_right}
    }
    \caption{Optimal solution for the 2-tone jamming waveform}
    \label{fig:absolute diff with the optimal solution}
\end{figure}

First, the convexity of the Lagrangian function $\mathcal{L}(\boldsymbol{\alpha},\boldsymbol{\omega},\lambda)$ with respect to $\omega_k$ is studied. In particular, this study is led for the optimal amplitude configuration $\boldsymbol{\alpha}^{\dagger}=\left[\sqrt{J/2},\sqrt{J/2}\right]^T$, for which $\mathbf{J}(\boldsymbol{\alpha}^{\dagger})=(J/2)\mathbf{I}_2$. Thus, from \eqref{eqn : C_espsilon star}, \eqref{eqn : partial derivative of the loss function with respect to omega_i} and \eqref{eqn : partial derivative of gamma matrix with respect to omega_i}, partial derivatives of ${\partial \mathcal{L}(\boldsymbol{\alpha}^{\dagger},\boldsymbol{\omega},\lambda^{\dagger}})$ with respect to $\omega_k$ are expressed as,
\begin{align}
\label{eqn : appendix derivee partielle de L par rapport à omega_1 pour puissance opt}
    \frac{\partial \mathcal{L}(\boldsymbol{\alpha}^{\dagger},\boldsymbol{\omega},\lambda^{\dagger})}{\partial\omega_1}&=(S+\sigma_n^2)^{-1}\frac{\partial D_{L/2}\left(\Delta_{21}\right)}{\partial \omega_1}
    \boldsymbol{1}_2^T\begin{bmatrix}
        c_1&c_2\\c_2&c_1
    \end{bmatrix}\times\nonumber\\
    &\qquad\qquad\qquad\begin{bmatrix}
        0&1\\1&0
    \end{bmatrix}
    \begin{bmatrix}
        c_1&c_2\\c_2&c_1
    \end{bmatrix}\boldsymbol{1}_2,\nonumber\\
    &=\frac{2}{S+\sigma_n^2}\frac{\partial D_{L/2}\left(\Delta_{21}\right)}{\partial \omega_1}(c_1+c_2)^2,
\end{align}
and,
\begin{equation}
\label{eqn : appendix derivee partielle de L par rapport à omega_2 pour puissance opt}
    \frac{\partial \mathcal{L}(\boldsymbol{\alpha}^{\dagger},\boldsymbol{\omega},\lambda^{\dagger})}{\partial\omega_2}=-\frac{2}{S+\sigma_n^2}\frac{\partial D_{L/2}\left(\Delta_{21}\right)}{\partial \omega_1}(c_1+c_2)^2,
\end{equation}
with,
\begin{align}
\label{eqn : c elements}
    \left\{ 
    \begin{array}{l}
    c_1=\nu(J/2,J/2)({J}/{2})(1+({J}/{2})L/(S+\sigma_n^2))\\
    c_2=-\nu(J/2,J/2)({J}/{2})^2(D_{L/2}(\Delta_{21})-1)/(S+\sigma_n^2)\\
    \nu(J/2,J/2)=\left[1+({J}/{2})\gamma/(S+\sigma_n^2)\right]^{-1}\times\\\qquad\left[1+({J}/{2})(L+D_{L/2}(\Delta_{21})-1)/(S+\sigma_n^2)\right]^{-1}
    \end{array}
    \right. .
\end{align}
From \eqref{eqn : partial derivative of the Dirichlet kernel}, \eqref{eqn : appendix derivee partielle de L par rapport à omega_1 pour puissance opt} and \eqref{eqn : appendix derivee partielle de L par rapport à omega_2 pour puissance opt}, ${\partial \mathcal{L}(\boldsymbol{\alpha}^{\dagger},\boldsymbol{\omega},\lambda^{\dagger}})/{\partial\omega_1}$ (resp. ${\partial \mathcal{L}(\boldsymbol{\alpha}^{\dagger},\boldsymbol{\omega},\lambda^{\dagger}})/{\partial\omega_2}$) is a non-monotonic function with respect to $\omega_1$ (resp. $\omega_2$) and $\mathcal{L}(\boldsymbol{\alpha}^{\dagger}, \boldsymbol{\omega},\lambda^{\dagger})$ is non-convex. In order to find the global minima of this function, its analytical form is first derived from \eqref{eqn : BMSE i est star} and \eqref{eqn : c elements}, 
\begin{align}
    &\mathcal{L}(\boldsymbol{\alpha}^{\dagger},\boldsymbol{\omega},\lambda^{\dagger}) = -J\nu(J/2,J/2)\left[1+({J}/{2})\gamma/(S+\sigma_n^2)\right],\nonumber\\
    &=-{(S+\sigma_n^2)}{\left[\frac{S+\sigma_n^2}{J}+\frac{L-1}{2}+\frac{D_{L/2}(\Delta_{21})}{2}\right]^{-1}}. 
\end{align}
Therefore, the Lagrangian function $\mathcal{L}(\boldsymbol{\alpha}^{\dagger},\boldsymbol{\omega},\lambda^{\dagger})$ is minimized for the $\Delta_{21}^{\star}$ value such that,
\begin{equation}
\label{eqn : pb opt}
\Delta_{21}^{\star}=\underset{\Delta_{21}}{\arg \min} \left(D_{L/2}(\Delta_{21})\right).
\end{equation}
A rough approximate solution to this problem is first found. By noticing that $D_{L/2}(\Delta_{21})$ is the product of $\sin(\frac{L+1}{2}\Delta_{21})$ and $1/\sin(\frac{1}{2}\Delta_{21})$, the latter being a decreasing function of $\Delta_{21}\in\left]0,\pi\right]$, the solution to \eqref{eqn : pb opt} can be approximated by the smallest value $\Delta_{21}$ that minimizes $\sin(\frac{L+1}{2}\Delta_{21})$. This first solution, denoted $\Delta_{21}^{\dagger(1)}$, is expressed as,
\begin{equation}
\label{eqn : delta_21 solution 1}
    \Delta_{21}^{\dagger(1)}={3\pi}/{(L+1)},
\end{equation}
and is depicted in Figure \ref{fig: Delta_21 approx 1} with the optimal solution $\Delta_{21}^{\star}$. From \eqref{eqn : delta_21 solution 1}, it is clear that the more $L$ increases, the smaller $\Delta_{21}^{\dagger(1)}$ is. From this observation, the function $1/\sin(\Delta_{21}/2)$ can be accurately approximated by its Taylor expansion $2/\Delta_{21}+o(1/\Delta_{21})$. Therefore, the problem \eqref{eqn : pb opt} is rewritten,    
\begin{equation}
\label{eqn : pb opt simplifié}
{\Delta}_{21}^{\dagger}=\underset{\Delta_{21}}{\arg \min} \left(\tilde{D}_{L/2}(\Delta_{21})\right),
\end{equation}
where,
\begin{equation}
\tilde{D}_{L/2}(\Delta_{21}) = (L+1)\times\textrm{sinc}\left(x\right),
\end{equation}
with $\textrm{sinc}(x)=\sin(x)/x$ and $x=\frac{L+1}{2}\times\Delta_{21}$, is a good approximate of $D_{L/2}(\Delta_{21})$ for high $L$ values. By the same reasoning as for $\Delta_{21}^{\dagger(1)}$, the solution to \eqref{eqn : pb opt simplifié} is obtained for the smallest value $\Delta_{21}$, denoted $\Delta_{21}^{\dagger(2)}$, such that ${\partial\tilde{D}_{L/2}(\Delta_{21})}/{\partial\omega_1}=0$. Thus, the partial derivative of $\tilde{D}_{L/2}(\Delta_{21})$ with respect to $\omega_1$ is first determined,
\begin{equation}
    \frac{\partial\tilde{D}_{L/2}(\Delta_{21})}{\partial\omega_1}=\frac{-\frac{L+1}{2}\cos(\frac{L+1}{2}\Delta_{21})+\sin(\frac{L+1}{2}\Delta_{21})}{\Delta_{21}^2}. 
\end{equation}
By equating this partial derivative to $0$, the equation to solve takes the form $x=\tan(x)$. The first solution to equation $x=\tan(x)$, which is the solution of interest as previously explained, is approximated by $x=4.493$ \cite{atlas_of_functions}[eq.(34:7:6)]. Consequently, the second approximate solution to problem \eqref{eqn : pb opt} is written,  
\begin{equation}
    \Delta_{21}^{\dagger(2)}={9}/{(L+1)}. 
\end{equation}
In order to choose the best approximation, the absolute difference $\epsilon_{\Delta_{21}^{\star}}=|\Delta_{21}^{\star}-\Delta_{21}^{\dagger(X)}|$ is calculated and depicted in Figure \ref{fig:absolute diff with the optimal solution} for $L$ ranging from $4$ to $128$. It is observed that both solutions converge to the optimal solution and that a faster convergence is obtained for $\Delta_{21}^{\dagger(2)}$. Consequently, $\Delta_{21}^{\dagger(2)}$ is preferred to $\Delta_{21}^{\dagger(1)}$. Finally, reminding that the Dirichlet kernel is an even $2\pi$-periodic function, the expressions \eqref{eqn : omega dagger} and \eqref{eqn : Delta_21 dagger} are obtained, ending the proof of Proposition \ref{Prop : 2-tone jamming signal}.

\subsection{BMSE calculation for the optimal jammer composed of $K=(L/2)+1$ tones}
\label{appendix : BMSE calculation for the optimal jammer composed of K=(L/2)+1 tones}
\renewcommand{\theequation}{E.\arabic{equation}}
\setcounter{equation}{0}
In this section, the analytical expression of the BMSE is determined for the jamming configuration $\{\boldsymbol{\alpha}^{\ddagger}, \boldsymbol{\omega}^{\ddagger},\}$ characterized by \eqref{eqn : alpha vector K L/2 + 1 tones} and \eqref{eqn : omega vector K L/2 + 1 tones}. For this configuration, it is first observed that $\left[\boldsymbol{\Gamma}(\boldsymbol{\omega})\right]_{(k+1)(k'+1)}=\left[\boldsymbol{\Gamma}(\boldsymbol{\omega})\right]_{kk'}$ and $\left[\boldsymbol{\Gamma}(\boldsymbol{\omega})\right]_{kK}=\left[\boldsymbol{\Gamma}(\boldsymbol{\omega})\right]_{(k+1)1}$. Therefore, $\boldsymbol{\Gamma}(\boldsymbol{\omega})$ is a circulant matrix that can be diagonalized in such a manner that $\boldsymbol{\Gamma}(\boldsymbol{\omega})=\mathbf{U}\boldsymbol{\Lambda}\mathbf{U}^H$, where $\mathbf{U}$ is an IDFT matrix and $\boldsymbol{\Lambda}$ is a diagonal matrix composed of the eigenvalues $\lambda_i=\sum_{k'=1}^{K}{\left[\boldsymbol{\Gamma}(\boldsymbol{\omega})\right]_{1k'}}\times\omega^{(i-1)(k'-1)}$, with $i\in\{1,\ldots,K\}$ and $\omega=e^{j\frac{2\pi}{K}}$. From \eqref{eqn : BMSE i est star} the BMSE of the interference estimate $\hat{i}^{\star}\left(\boldsymbol{\alpha}^{\ddagger},\boldsymbol{\omega}^{\ddagger}\right)$ is then written,
\begin{align}
    \label{eqn : appendix BMSE ddagger}&\mathrm{BMSE}\left(\,\hat{i}^{\star}\left(\boldsymbol{\alpha}^{\ddagger},\boldsymbol{\omega}^{\ddagger}\right)\,\right)=\boldsymbol{1}_K^T\left(\frac{K}{J}\mathbf{I}_K+\frac{\mathbf{U}\boldsymbol{\Lambda}\mathbf{U}^H}{S+\sigma_n^2}\right)^{-1}\boldsymbol{1}_K,\nonumber\\
    &=\boldsymbol{1}_K^T\mathbf{U}\left(\frac{K}{J}\mathbf{I}_K+\frac{\boldsymbol{\Lambda}}{S+\sigma_n^2}\right)^{-1}\mathbf{U}^H\boldsymbol{1}_K,\nonumber\\
    &=K\left[ 1,0,\ldots,0\right]
    \mathrm{diag}\left(\zeta_1,...,\zeta_{K}\right)
    \left[1,0,\ldots,0\right]^T,\nonumber\\
    &=K\zeta_1, 
\end{align}
where $\zeta_k=\left({K}/{J}+{\lambda_k}/{(S+\sigma_n^2)}\right)^{-1}$.
Therefore, the BMSE highly depends on the eigenvalue $\lambda_1$, which is expressed as,
\begin{align}
    \lambda_1 &= \sum_{k'=1}^{K}{\left[\boldsymbol{\Gamma}(\boldsymbol{\omega})\right]_{1k'}},\nonumber\\
    &=L-K+1+\sum_{k'=2}^{K}{D_{L/2}\left(\frac{2\pi}{K}(k'-1)\right)},\nonumber\\
    &=L-K+1+\sum_{l=-L/2}^{l=L/2}{\sum_{k'=2}^{K}{e^{jl\frac{2\pi}{K}(k'-1)}}},\nonumber\\
    &=L+\sum_{\underset{l\neq 0}{l=-L/2}}^{L/2}{f(l)},
\end{align}
where, 
\begin{equation}
    \label{eqn : f(l)}f(l)=\frac{\cos{(l\pi)}\sin{\left({l\pi}(K-1)/K\right)}}{\sin{\left({l\pi}/{K}\right)}}.
\end{equation}
It can be shown that the function $f(l)$ admits two values. For $l$ values such that $l=nK$, with $n\in\mathbb{N}^{*}$, l'Hôpital's rule is used to obtain $f(l)=K-1$. The number of values $l$ that respect this condition is equal to $2\floor*{\frac{L}{2K}}$. For other values of $l$, 
$f(l)=-1$ is obtained from \eqref{eqn : f(l)} using the angle sum identity. Consequently, $\lambda_1$ simplifies to,
\begin{align}
\label{eqn : lambda_0}
    \lambda_1 &= L + (K-1) \times2\floor*{\frac{L}{2K}} -1\times\left(L-2\floor*{\frac{L}{2K}}\right),\nonumber\\
    &=2K\floor*{\frac{L}{2K}}.
\end{align}
Substituting \eqref{eqn : lambda_0} into \eqref{eqn : appendix BMSE ddagger}, \eqref{eqn : BMSE ddagger} is obtained, ending the proof of Proposition \ref{Prop : (L/2+1)-tone jamming signal}.

\balance
\bibliographystyle{IEEEtran} 
\bibliography{ref.bib}

\begin{thebibliography}{10}
\providecommand{\url}[1]{#1}
\csname url@samestyle\endcsname
\providecommand{\newblock}{\relax}
\providecommand{\bibinfo}[2]{#2}
\providecommand{\BIBentrySTDinterwordspacing}{\spaceskip=0pt\relax}
\providecommand{\BIBentryALTinterwordstretchfactor}{4}
\providecommand{\BIBentryALTinterwordspacing}{\spaceskip=\fontdimen2\font plus
\BIBentryALTinterwordstretchfactor\fontdimen3\font minus
  \fontdimen4\font\relax}
\providecommand{\BIBforeignlanguage}[2]{{%
\expandafter\ifx\csname l@#1\endcsname\relax
\typeout{** WARNING: IEEEtran.bst: No hyphenation pattern has been}%
\typeout{** loaded for the language `#1'. Using the pattern for}%
\typeout{** the default language instead.}%
\else
\language=\csname l@#1\endcsname
\fi
#2}}
\providecommand{\BIBdecl}{\relax}
\BIBdecl

\bibitem{liu_physical_2017}
\BIBentryALTinterwordspacing
Y.~Liu, H.-H. Chen, and L.~Wang, ``Physical {Layer} {Security} for {Next}
  {Generation} {Wireless} {Networks}: {Theories}, {Technologies}, and
  {Challenges},'' \emph{IEEE Communications Surveys \& Tutorials}, vol.~19,
  no.~1, pp. 347--376, 2017. [Online]. Available:
  \url{http://ieeexplore.ieee.org/document/7539590/}
\BIBentrySTDinterwordspacing

\bibitem{yue_low_2023}
\BIBentryALTinterwordspacing
P.~Yue, J.~An, J.~Zhang, J.~Ye, G.~Pan, S.~Wang, P.~Xiao, and L.~Hanzo, ``Low
  {Earth} {Orbit} {Satellite} {Security} and {Reliability}: {Issues},
  {Solutions}, and the {Road} {Ahead},'' \emph{IEEE Communications Surveys \&
  Tutorials}, vol.~25, no.~3, pp. 1604--1652, 2023. [Online]. Available:
  \url{https://ieeexplore.ieee.org/document/10209551/}
\BIBentrySTDinterwordspacing

\bibitem{tusha_interference_2025}
\BIBentryALTinterwordspacing
A.~Tusha and H.~Arslan, ``Interference {Burden} in {Wireless} {Communications}:
  {A} {Comprehensive} {Survey} {From} {PHY} {Layer} {Perspective},'' \emph{IEEE
  Communications Surveys \& Tutorials}, vol.~27, no.~4, pp. 2204--2246, Aug.
  2025. [Online]. Available:
  \url{https://ieeexplore.ieee.org/document/10736552/}
\BIBentrySTDinterwordspacing

\bibitem{morales-ferre_survey_2020}
\BIBentryALTinterwordspacing
R.~Morales-Ferre, P.~Richter, E.~Falletti, A.~De~La~Fuente, and E.~S. Lohan,
  ``A {Survey} on {Coping} {With} {Intentional} {Interference} in {Satellite}
  {Navigation} for {Manned} and {Unmanned} {Aircraft},'' \emph{IEEE
  Communications Surveys \& Tutorials}, vol.~22, no.~1, pp. 249--291, 2020.
  [Online]. Available: \url{https://ieeexplore.ieee.org/document/8882350/}
\BIBentrySTDinterwordspacing

\bibitem{hegarty_suppression_2000}
C.~Hegarty, A.~J. Van~Dierendonck, D.~Bobyn, M.~Tran, and J.~Grabowski,
  ``\BIBforeignlanguage{en}{Suppression of {Pulsed} {Interference} through
  {Blanking}},'' Jun. 2000, pp. 399--408.

\bibitem{borio_gnss_2012}
\BIBentryALTinterwordspacing
D.~Borio, C.~O'Driscoll, and J.~Fortuny, ``{GNSS} {Jammers}: {Effects} and
  countermeasures,'' in \emph{2012 6th {ESA} {Workshop} on {Satellite}
  {Navigation} {Technologies} ({Navitec} 2012) \& {European} {Workshop} on
  {GNSS} {Signals} and {Signal} {Processing}}, Dec. 2012, pp. 1--7, iSSN:
  2325-5455. [Online]. Available:
  \url{https://ieeexplore.ieee.org/document/6423048/}
\BIBentrySTDinterwordspacing

\bibitem{borio_swept_2016}
\BIBentryALTinterwordspacing
D.~Borio, ``Swept {GNSS} jamming mitigation through pulse blanking,'' in
  \emph{2016 {European} {Navigation} {Conference} ({ENC})}.\hskip 1em plus
  0.5em minus 0.4em\relax Helsinki, Finland: IEEE, May 2016, pp. 1--8.
  [Online]. Available: \url{http://ieeexplore.ieee.org/document/7530549/}
\BIBentrySTDinterwordspacing

\bibitem{aygur_narrowband_2025}
\BIBentryALTinterwordspacing
M.~Aygur, S.~Kandeepan, A.~Giorgetti, A.~Al-Hourani, E.~Arbon, and M.~Bowyer,
  ``Narrowband {Interference} {Mitigation} {Techniques}: {A} {Survey},''
  \emph{IEEE Communications Surveys \& Tutorials}, pp. 1--1, 2025. [Online].
  Available: \url{https://ieeexplore.ieee.org/document/10845757/}
\BIBentrySTDinterwordspacing

\bibitem{masry_closed-form_1984}
\BIBentryALTinterwordspacing
E.~Masry, ``\BIBforeignlanguage{en}{Closed-{Form} {Analytical} {Results} for
  the {Rejection} of {Narrow}-{Band} {Interference} in {PN} {Spread}-{Spectrum}
  {Systems}--{Part} {I}: {Linear} {Prediction} {Filters}},''
  \emph{\BIBforeignlanguage{en}{IEEE Transactions on Communications}}, vol.~32,
  no.~8, pp. 888--896, Aug. 1984. [Online]. Available:
  \url{http://ieeexplore.ieee.org/document/1096164/}
\BIBentrySTDinterwordspacing

\bibitem{wang_rejection_1988}
\BIBentryALTinterwordspacing
Y.-C. Wang and L.~Milstein, ``Rejection of multiple narrow-band interference in
  both {BPSK} and {QPSK} {DS} spread-spectrum systems,'' \emph{IEEE
  Transactions on Communications}, vol.~36, no.~2, pp. 195--204, Feb. 1988.
  [Online]. Available: \url{https://ieeexplore.ieee.org/document/2750/}
\BIBentrySTDinterwordspacing

\bibitem{ketchum_adaptive_1982}
\BIBentryALTinterwordspacing
J.~Ketchum and J.~Proakis, ``Adaptive {Algorithms} for {Estimating} and
  {Suppressing} {Narrow}-{Band} {Interference} in {PN} {Spread}-{Spectrum}
  {Systems},'' \emph{IEEE Transactions on Communications}, vol.~30, no.~5, pp.
  913--924, May 1982. [Online]. Available:
  \url{http://ieeexplore.ieee.org/document/1095542/}
\BIBentrySTDinterwordspacing

\bibitem{young_analysis_1998}
\BIBentryALTinterwordspacing
J.~Young and J.~Lehnert, ``Analysis of {DFT}-based frequency excision
  algorithms for direct-sequence spread-spectrum communications,'' \emph{IEEE
  Transactions on Communications}, vol.~46, no.~8, pp. 1076--1087, Aug. 1998.
  [Online]. Available: \url{http://ieeexplore.ieee.org/document/705409/}
\BIBentrySTDinterwordspacing

\bibitem{krim_two_1996}
\BIBentryALTinterwordspacing
H.~Krim and M.~Viberg, ``Two decades of array signal processing research: the
  parametric approach,'' \emph{IEEE Signal Processing Magazine}, vol.~13,
  no.~4, pp. 67--94, Jul. 1996. [Online]. Available:
  \url{http://ieeexplore.ieee.org/document/526899/}
\BIBentrySTDinterwordspacing

\bibitem{myrick_low_2001}
\BIBentryALTinterwordspacing
W.~Myrick, J.~Goldstein, and M.~Zoltowski, ``Low complexity anti-jam space-time
  processing for {GPS},'' in \emph{2001 {IEEE} {International} {Conference} on
  {Acoustics}, {Speech}, and {Signal} {Processing}. {Proceedings} ({Cat}.
  {No}.{01CH37221})}.\hskip 1em plus 0.5em minus 0.4em\relax Salt Lake City,
  UT, USA: IEEE, 2001, pp. 2233--2236. [Online]. Available:
  \url{http://ieeexplore.ieee.org/document/940442/}
\BIBentrySTDinterwordspacing

\bibitem{musumeci_use_2014}
\BIBentryALTinterwordspacing
L.~Musumeci and F.~Dovis, ``\BIBforeignlanguage{en}{Use of the {Wavelet}
  {Transform} for {Interference} {Detection} and {Mitigation} in {Global}
  {Navigation} {Satellite} {Systems}},''
  \emph{\BIBforeignlanguage{en}{International Journal of Navigation and
  Observation}}, vol. 2014, pp. 1--14, Feb. 2014. [Online]. Available:
  \url{https://www.hindawi.com/journals/ijno/2014/262186/}
\BIBentrySTDinterwordspacing

\bibitem{hu_narrowband_2025}
\BIBentryALTinterwordspacing
Y.~Hu, S.~Huang, L.~Zhao, and M.~Jiang, ``Narrowband {Interference}
  {Cancellation} for {OFDM} {Based} on {Deep} {Learning} and {Compressed}
  {Sensing},'' \emph{IEEE Transactions on Signal Processing}, vol.~73, pp.
  1612--1625, 2025. [Online]. Available:
  \url{https://ieeexplore.ieee.org/document/10772564/}
\BIBentrySTDinterwordspacing

\bibitem{huang_reweighted_2019}
\BIBentryALTinterwordspacing
Y.~Huang, G.~Liao, Y.~Xiang, Z.~Zhang, J.~Li, and A.~Nehorai, ``Reweighted
  {Nuclear} {Norm} and {Reweighted} {Frobenius} {Norm} {Minimizations} for
  {Narrowband} {RFI} {Suppression} on {SAR} {System},'' \emph{IEEE Transactions
  on Geoscience and Remote Sensing}, vol.~57, no.~8, pp. 5949--5962, Aug. 2019.
  [Online]. Available: \url{https://ieeexplore.ieee.org/document/8681711/}
\BIBentrySTDinterwordspacing

\bibitem{borio_two-pole_2008}
\BIBentryALTinterwordspacing
D.~Borio, L.~Camoriano, and L.~Lo~Presti, ``Two-{Pole} and {Multi}-{Pole}
  {Notch} {Filters}: {A} {Computationally} {Effective} {Solution} for {GNSS}
  {Interference} {Detection} and {Mitigation},'' \emph{IEEE Systems Journal},
  vol.~2, no.~1, pp. 38--47, Mar. 2008. [Online]. Available:
  \url{http://ieeexplore.ieee.org/document/4433998/}
\BIBentrySTDinterwordspacing

\bibitem{chien_design_2015}
\BIBentryALTinterwordspacing
Y.-R. Chien, ``Design of {GPS} {Anti}-{Jamming} {Systems} {Using} {Adaptive}
  {Notch} {Filters},'' \emph{IEEE Systems Journal}, vol.~9, no.~2, pp.
  451--460, Jun. 2015. [Online]. Available:
  \url{https://ieeexplore.ieee.org/document/6631513/}
\BIBentrySTDinterwordspacing

\bibitem{masry_closed-form_1985}
\BIBentryALTinterwordspacing
E.~Masry, ``\BIBforeignlanguage{en}{Closed-{Form} {Analytical} {Results} for
  the {Rejection} of {Narrow}-{Band} {Interference} in {PN} {Spread}-{Spectrum}
  {Systems}--{Part} {II}: {Linear} {Interpolation} {Filters}},''
  \emph{\BIBforeignlanguage{en}{IEEE Transactions on Communications}}, vol.~33,
  no.~1, pp. 10--19, Jan. 1985. [Online]. Available:
  \url{http://ieeexplore.ieee.org/document/1096204/}
\BIBentrySTDinterwordspacing

\bibitem{loh-ming_li_rejection_1982}
\BIBentryALTinterwordspacing
{Loh-Ming Li} and L.~Milstein, ``Rejection of {Narrow}-{Band} {Interference} in
  {PN} {Spread}-{Spectrum} {Systems} {Using} {Transversal} {Filters},''
  \emph{IEEE Transactions on Communications}, vol.~30, no.~5, pp. 925--928, May
  1982. [Online]. Available: \url{http://ieeexplore.ieee.org/document/1095543/}
\BIBentrySTDinterwordspacing

\bibitem{iltis_performance_1984}
\BIBentryALTinterwordspacing
R.~Iltis and L.~Milstein, ``\BIBforeignlanguage{en}{Performance {Analysis} of
  {Narrow}-{Band} {Interference} {Rejection} {Techniques} in {DS}
  {Spread}-{Spectrum} {Systems}},'' \emph{\BIBforeignlanguage{en}{IEEE
  Transactions on Communications}}, vol.~32, no.~11, pp. 1169--1177, Nov. 1984.
  [Online]. Available: \url{http://ieeexplore.ieee.org/document/1095986/}
\BIBentrySTDinterwordspacing

\bibitem{masry_performance_1986}
\BIBentryALTinterwordspacing
E.~Masry and L.~Milstein, ``Performance of {DS} {Spread}-{Spectrum} {Receiver}
  {Employing} {Interference}-{Suppression} {Filters} {Under} a {Worst}-{Case}
  {Jamming} {Condition},'' \emph{IEEE Transactions on Communications}, vol.~34,
  no.~1, pp. 13--21, 1986. [Online]. Available:
  \url{http://ieeexplore.ieee.org/document/1096425/}
\BIBentrySTDinterwordspacing

\bibitem{fonteneau_rejection_2024}
\BIBentryALTinterwordspacing
C.~Fonteneau, M.~Crussière, A.~Bazin, and O.~P. Pasquero, ``Rejection
  {Capability} of {Anti}-{Jamming} {Wiener} {Filter} for {Multi}-{Tone}
  {Interference} in {DSSS} {Systems},'' in \emph{{MILCOM} 2024 - 2024 {IEEE}
  {Military} {Communications} {Conference} ({MILCOM})}.\hskip 1em plus 0.5em
  minus 0.4em\relax Washington, DC, USA: IEEE, Oct. 2024, pp. 1--6. [Online].
  Available: \url{https://ieeexplore.ieee.org/document/10773998/}
\BIBentrySTDinterwordspacing

\bibitem{Funda_Stat_Signal_Processing_EstTheory}
S.~M. Kay, \emph{Fundamentals of statistical signal processing: estimation
  theory}.\hskip 1em plus 0.5em minus 0.4em\relax USA: Prentice-Hall, Inc.,
  1993.

\bibitem{goldstein_multistage_1998}
\BIBentryALTinterwordspacing
J.~Goldstein, I.~Reed, and L.~Scharf, ``A multistage representation of the
  {Wiener} filter based on orthogonal projections,'' \emph{IEEE Transactions on
  Information Theory}, vol.~44, no.~7, pp. 2943--2959, Nov. 1998. [Online].
  Available: \url{http://ieeexplore.ieee.org/document/737524/}
\BIBentrySTDinterwordspacing

\bibitem{honig_adaptive_2002}
\BIBentryALTinterwordspacing
M.~Honig and J.~Goldstein, ``\BIBforeignlanguage{en}{Adaptive reduced-rank
  interference suppression based on the multistage {Wiener} filter},''
  \emph{\BIBforeignlanguage{en}{IEEE Transactions on Communications}}, vol.~50,
  no.~6, pp. 986--994, Jun. 2002. [Online]. Available:
  \url{http://ieeexplore.ieee.org/document/1010618/}
\BIBentrySTDinterwordspacing

\bibitem{song_adaptive_2012}
\BIBentryALTinterwordspacing
N.~Song, R.~C. De~Lamare, M.~Haardt, and M.~Wolf, ``Adaptive {Widely} {Linear}
  {Reduced}-{Rank} {Interference} {Suppression} {Based} on the {Multistage}
  {Wiener} {Filter},'' \emph{IEEE Transactions on Signal Processing}, vol.~60,
  no.~8, pp. 4003--4016, Aug. 2012. [Online]. Available:
  \url{https://ieeexplore.ieee.org/document/6194364/}
\BIBentrySTDinterwordspacing

\bibitem{zhang_fast_2016}
\BIBentryALTinterwordspacing
M.~Zhang, A.~Zhang, and J.~Li, ``Fast and {Accurate} {Rank} {Selection}
  {Methods} for {Multistage} {Wiener} {Filter},'' \emph{IEEE Transactions on
  Signal Processing}, vol.~64, no.~4, pp. 973--984, Feb. 2016. [Online].
  Available: \url{http://ieeexplore.ieee.org/document/7305823/}
\BIBentrySTDinterwordspacing

\bibitem{kingma_adam_2017}
\BIBentryALTinterwordspacing
D.~P. Kingma and J.~Ba, ``Adam: {A} {Method} for {Stochastic} {Optimization},''
  Jan. 2017, arXiv:1412.6980. [Online]. Available:
  \url{http://arxiv.org/abs/1412.6980}
\BIBentrySTDinterwordspacing

\bibitem{Minka2000OldAN}
\BIBentryALTinterwordspacing
T.~P. Minka, ``Old and new matrix algebra useful for statistics,'' 2000.
  [Online]. Available: \url{https://api.semanticscholar.org/CorpusID:15971655}
\BIBentrySTDinterwordspacing

\bibitem{Petersen2006TheMC}
\BIBentryALTinterwordspacing
K.~B. Petersen and M.~S. Pedersen, ``The matrix cookbook,'' 2006. [Online].
  Available: \url{https://api.semanticscholar.org/CorpusID:1221763}
\BIBentrySTDinterwordspacing

\bibitem{atlas_of_functions}
K.~Oldham and Spanier, \emph{Atlas of functions: with equator, the atlas
  function calculator/ by Keith Oldham, Jan Myland and Jerome Spanier},
  2nd~ed.\hskip 1em plus 0.5em minus 0.4em\relax New York: Springer, 2009.

\end{thebibliography}

\vspace{12pt}

\end{document}